\documentclass[10pt, a4paper,aps,prl,superscriptaddress,floatfix,twocolumn]{revtex4-2}

\usepackage[english]{babel}
\usepackage[utf8]{inputenc}
\usepackage{csquotes}
\usepackage{color}
\usepackage{enumitem}
\usepackage[section]{placeins}
\usepackage{setspace}

\usepackage{amssymb}
\usepackage{mathrsfs}  
\usepackage{amsthm}
\usepackage{amsmath}
\usepackage{mathtools}
\usepackage{bigints}
\usepackage{esint}
\usepackage{siunitx}

\usepackage{hyperref}
\usepackage[final]{todonotes}
\usepackage{lipsum}
\usepackage{layouts}
\usepackage{nomencl}

\usepackage{svg}
\svgpath{{graphics/}}
\usepackage{graphicx}
\graphicspath{{graphics/}}
\usepackage[section]{placeins}
\usepackage{bibunits}

\newcommand{\F}{F}                              
\newcommand{\FCA}{\F_{CH}}                      
\newcommand{\FLC}{\F_{EC}}                      
\newcommand{\FIN}{\F_{IN}}                      

\newcommand{\dt}{\partial_t}                    

\newcommand{\eps}{\epsilon}                     
\newcommand{\CA}{Ca}                 
\newcommand{\CAF}{\frac{1}{\CA}}                 
\newcommand{\LCF}{Ec}                 
\newcommand{\IN}{In}                 
\newcommand{\INF}{\frac{1}{\IN}}                 

\newcommand{\surf}{\mathcal{S}}                 
\newcommand{\normal}{\boldsymbol{\nu}}                         
\newcommand{\gradS}{\nabla_\surf}               
\newcommand{\guentGrad}{\nabla_C}               
\newcommand{\tangGrad}{\nabla_P}               
\newcommand{\laplaceS}{\Delta_\surf}               
\newcommand{\surfint}[1]{\int_\surf #1\,d\surf} 
\newcommand{\princCurv}{k}                      
\newcommand{\curvTensor}{\boldsymbol{B}}                       

\newcommand{\R}{\mathbb{R}}                     
\newcommand{\func}{\phi}                        

\newcommand{\rCyl}{r_{Cyl}}                     

\begin{document}
\begin{bibunit}[plainnat]

\title{Coordinated motion of epithelial layers on curved surfaces}

\author{L. Happel}
\affiliation{Institute of Scientific Computing, TU Dresden, 01062 Dresden, Germany}

\author{A. Voigt}
\affiliation{Institute of Scientific Computing, TU Dresden, 01062 Dresden, Germany}
\affiliation{Center for Systems Biology Dresden, Pfotenhauerstr. 108, 01307 Dresden, Germany}
\affiliation{Cluster of Excellence, Physics of Life, TU Dresden, Arnoldstr. 18, 01307 Dresden, Germany}

\begin{abstract}
Coordinated cellular movements are key processes in tissue morphogenesis. Using a cell-based modeling approach we study the dynamics of epithelial layers lining surfaces with constant and varying curvature. We demonstrate that extrinsic curvature effects can explain the alignment of cell elongation with the principal directions of curvature. Together with specific self-propulsion mechanisms and cell-cell interactions this effect gets enhanced and can explain observed large-scale, persistent and circumferential rotation on cylindrical surfaces. On toroidal surfaces the resulting curvature coupling is an interplay of intrinsic and extrinsic curvature effects. These findings unveil the role of curvature and postulate its importance for tissue morphogenesis. 
\end{abstract}

\maketitle

Geometry, and in particular local curvature, influences biological systems at various length scales \cite{Schamberger_AM_2022}. One example associated with curved epithelial layers is collective rotation. Persistent and synchronous rotation on a sphere has been observed in vivo \cite{Haigoetal_Science_2011,Satoetal_NComm_2015,Barlanetal_DC_2017,Fernandezetal_NP_2021,Tan_bioRxiv_2022}, in vitro \cite{Tanneretal_PNAS_2012,Wangetal_PNAS_2013,Chinetal_PNAS_2018,Hirataetal_CS_2018} and in silico \cite{Happel_EPL_2022,Tan_bioRxiv_2022}. These phenomena differ significantly from collective behavior in flat space and are attributed to the geometric and topological properties of the sphere. Nevertheless, the underlying mechanisms that trigger such collective rotation remain unclear even for surfaces as simple as a sphere, not to mention the curved environments that epithelial tissues encounter during morphogenesis. To better understand how curvature influences epithelial layers we consider two prototypical geometries which allow for validation for specific cell types \cite{Glentis_SA_2022,Yu_Biomaterials_2021}.

At the single cell level it has been shown that cells sense and respond to curvature \cite{Bade_BJ_2018,Pieuchot_NC_2018,Werner_AB_2019}, essentially by regulating the transcellular network architecture \cite{Baptistaetal_TB_2019,Callens_BM_2020,Harmandetal_PRX_2021} and aligning the filaments with the principal curvature directions \cite{Callens_BM_2020}. Experimental realizations furthermore show a dependence on cell type, while, e.g., filaments of fibroblasts align with the minimal curvature direction \cite{Bade_BJ_2018,Callens_BM_2020}, the elongation direction of MDCK cells aligns with the maximal curvature direction \cite{Callens_BM_2020}. Also the nucleus plays a role and cell migration on curved surfaces is shown to follow the path of least nuclear mechanical stress \cite{Werner_AB_2019,Pieuchot_NC_2018}. These phenomena, which describe the response to cell-scale curvature, are termed curvotaxis \cite{Pieuchot_NC_2018} and can be extended to collective cell behavior on curved surfaces. Coordinated rotation has been associated with the alignment of filaments with principal curvature directions, cell-cell adhesion and apical-basal polarity \cite{Tanneretal_PNAS_2012,Wangetal_PNAS_2013}. In \cite{Glentis_SA_2022} cylindrical epithelia of MDCK cells are considered. The results indicate that proper cell-cell adhesion is essential, as well as aligned cellular polar order. This alignment is again in the principal curvature directions. In contrast, the orientation of the actin network does not seem to be essential for collective rotation. Also geometries with varying curvature, e.g. toroidal surfaces have been considered \cite{Yu_Biomaterials_2021}. However, in \cite{Yu_Biomaterials_2021} only cell elongation is addressed. 

In this Letter we propose a minimal cell-based surface model that reproduces these effects for MDCK cells. Two-dimensional vertex models, e.g., \cite{Farhadifaretal_CB_2007,Bietal_PRX_2016,Popovicetal_NJP_2021} and multi-phase field models \cite{Nonomura_PLOS_2012,Camley_PNAS_2014,Loeber_SR_2015,Palmieri_SR_2015,Marth_IF_2016,Mueller_PRL_2019,Wenzel_PRE_2021} have been successfully used to simulate epithelial tissue in flat space.  Extensions to curved surfaces are still rare, see \cite{Sussman_PRR_2020,Tan_bioRxiv_2022} for vertex models and \cite{Happel_EPL_2022} for multi-phase field models considered on a sphere. None of these approaches account for extrinsic curvature contributions. These terms, which somehow translate the three-dimensional nature of a thin layer, for an epithelial layer, e.g. the difference between the apical and basal side, into an effectively two-dimensional framework on the curved surface, will be shown to be essential to model the curvature effects discussed above. Extrinsic curvature effects are well established in the theory of surface liquid crystals \cite{Napoli_PRE_2012,Napoli_PRL_2012}. These theories force the director field to be tangential to the surface and the corresponding free energies contain coupling terms between the director field and the principal curvature directions of the shape operator \cite{Napoli_PRE_2012,Napoli_PRL_2012,Nestler_JNS_2018,Nitschke_PRSA_2018}. These terms follow naturally if the energies are derived as thin film limits from three-dimensional theories \cite{Golovaty_JNS_2017,Nestler_JNS_2018,Nitschke_PRSA_2018} and have shown various implications on phase transitions \cite{Nestler_SM_2020}, active nematodynamic flows \cite{Napoli_PRE_2020,Nestler_CCP_2022,Bell_PRL_2022} and shape deformations \cite{Nitschke_PRSA_2020}.

\begin{figure}[htb]
    \centering
    \includegraphics[width=0.49\textwidth]{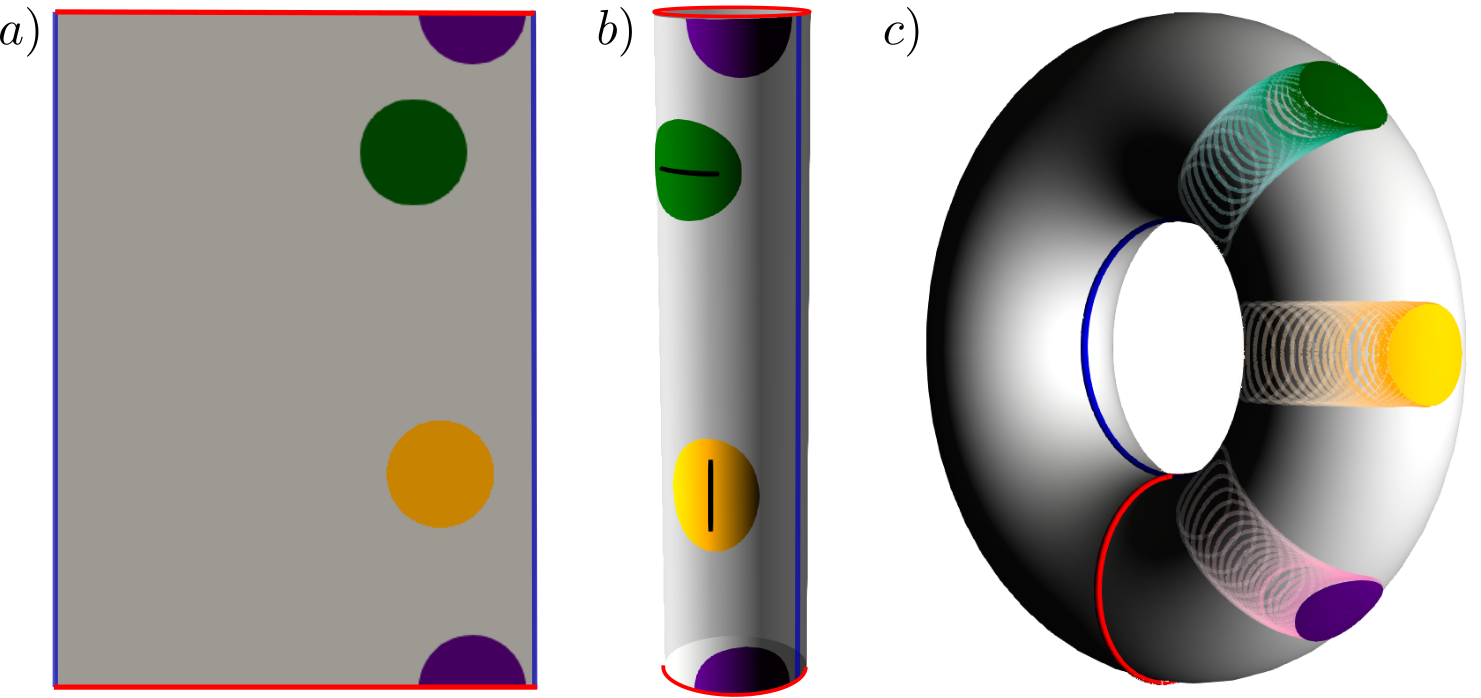} 
    \caption{Geometries and cell shapes. Red and blue lines mark periodic boundaries, which are glued together in b) and c) (not to scale). Three individual cells are shown in their equilibrium configuration in a) and b). The colors correspond to the parameter $\LCF$ which models extrinsic curvature effects, see eq. \eqref{eq:FEC}. $\LCF = 0$ (purple) leads to a (geodesic) circle on both geometries. $\LCF > 0$ (green) favours an alignment in direction of maximum absolute curvature and $\LCF < 0$ (yellow) an alignment in direction of minimal absolute curvature. The elongation is marked and enhanced for visibility. On toroidal surfaces cell shapes depend on position. In c) trajectories and final positions and shapes of the cells are shown. The effect of extrinsic curvature is not visible. All shapes are obtained by solving eq. \eqref{eq:evol} with $v_0 = 0$.}
    \label{fig_ext_curv}
\end{figure}

We consider a multi-phase field model that allows for cell deformations, local cellular rearrangements and detailed cell-cell interactions, as well as extrinsic curvature coupling \cite{Wenzel_PRE_2021,Jain_SR_2023}. We consider two-dimensional phase field variables $\func_i(\mathbf{x},t)$ one for each cell, with $\mathbf{x}$ defined on the surface ${\cal{S}}$. Values of $\func_i=1$ and $\func_i=-1$ denote the interior and exterior of a cell. The cell boundary is implicitly defined as the zero-level set of $\func_i$. We consider various topologically equivalent surfaces ${\cal{S}}$, see Figure \ref{fig_ext_curv}. 
The dynamics for each $\func_i$ reads
\begin{equation} \label{eq:evol}
    \dt \func_i + v_0(\mathbf{v}_i \cdot \gradS\func_i) = \laplaceS \frac{\delta \F}{\delta \func_i},
\end{equation}
for $i=1,...,N$, where $N$ denotes the number of cells. $\F$ is a free energy and $\mathbf{v}_i$ a vector field used to incorporate activity, with a self-propulsion strength $v_0$. The operators $\gradS$ and $\laplaceS$ denote the covariant derivative and Laplace-Beltrami operator on ${\cal{S}}$, respectively. All quantities are non-dimensional quantities. As in previous studies \cite{Happel_EPL_2022, Marth_JRSI_2015, Marth_IF_2016, Wenzel_JCP_2019, Wenzel_CMAM_2021, Wenzel_PRE_2021}, we consider conserved dynamics. 

The free energy reads
$\F = \FCA + \FLC + \FIN$. The first contribution is a (de Gennes-)Cahn-Hilliard energy \cite{Salvalaglio_MMAS_2021,Benes_Arxiv_2023}
\begin{equation*}
    \FCA = \sum\limits_{i=1}^N\CAF \surfint{ \frac{1}{G(\func_i)} \left(\frac{\eps}{2}\|\gradS\func_i\|^2 + \frac{1}{\eps}W(\func_i)\right)},
\end{equation*}
which stabilizes the interface, with $W(\phi_i) = \tfrac14 (1 - \func_i^2)^2$ a double-well potential, $\eps$ a small parameter determining the width of the diffuse interface and $1/G(\phi_i)$ a de Gennes coefficient. This term does not influence the asymptotic limit ($\eps \to 0$) \cite{Salvalaglio_MMAS_2021} but helps to keep $-1 \leq \phi_i \leq 1$, which becomes important on curved surfaces \cite{Benes_Arxiv_2023}. We consider $G(\phi_i) = \frac{3}{2}|1 - \phi_i^2|$. $\CA$ is the capillary number. This covariant formulation only accounts for intrinsic curvature effects. Minimizing this energy by solving eq. \eqref{eq:evol} with $v_0 = 0$ on a cylindrical surface leads to a geodesic circle with no preferred orientation, see Figure \ref{fig_ext_curv} (purple cell). This does not resample the observed properties of single cells \cite{Callens_BM_2020}. 
 
We associate a director field with the cell shape. In flat space this has been considered in \cite{Mueller_PRL_2019,Wenzel_PRE_2021}. Adapting the definition to the surface we obtain the surface Q-tensor fields
 \begin{equation*}
    \mathbf{q}_i=\begin{pmatrix} \surfint{\frac{(\partial_{\mathbf{t}_2}\func_i)^2-(\partial_{\mathbf{t}_1}\func_i)^2}{2}} & \surfint{-\partial_{\mathbf{t}_1}\func_i\partial_{\mathbf{t}_2}\func_i} \\[6pt]
    \surfint{-\partial_{\mathbf{t}_1}\func_i\partial_{\mathbf{t}_2}\func_i} & \surfint{\frac{(\partial_{\mathbf{t}_1}\func_i)^2-(\partial_{\mathbf{t}_2}\func_i)^2}{2}}
    \end{pmatrix}
\end{equation*}
where $\mathbf{t}_1(\mathbf{x})$ and $\mathbf{t}_2(\mathbf{x})$ denote orthonormal vectors of the tangent plane at $\mathbf{x} \in {\cal{S}}$ which are related by parallel transport to the principal curvature directions in the center of mass of the cell $i$, see SI for details, which includes Refs. \cite{Bai_JGT_2008,Crane_ACM_2013}. Together with $\normal(\mathbf{x})$, the outward-pointing normal to the surface ${\cal{S}}$, they define the Darboux frame. The eigenvectors of the tensor fields $\mathbf{q}_i$ correspond to the direction of largest elongation and contraction and the corresponding eigenvalues measure the degree of deformation. Using these directions to define director fields $\mathbf{d}_i$ allows to associate nematic order to the epithelial tissue \cite{Doostmohammadi_NC_2018,Mueller_PRL_2019,Wenzel_PRE_2021,Happel_EPL_2022}. In our case $\mathbf{q}_i$ and $\mathbf{d}_i$ are tangential tensor and vector fields, respectively. Coarse-grained quantities of the surface Q-tensor fields $\mathbf{q}$ and the director fields $\mathbf{d}$ are considered in surface liquid crystal models and related by $
\mathbf{q} = S \left(\mathbf{d} \otimes \mathbf{d} - \frac{1}{2} \mathbf{g} \right)
$ \cite{Nestler_SM_2020},
where $S$ is a nematic order parameter and $\mathbf{g}$ is the metric of the surface ${\cal{S}}$. Already in typical one-constant approximations of the corresponding surface energies, if derived as a thin film limit from the corresponding 3D models, additional geometric coupling terms occur \cite{Napoli_PRE_2012,Napoli_PRL_2012,Nestler_JNS_2018,Nitschke_PRSA_2018}. In case of the surface Frank-Oseen model the term of interest reads
\begin{equation} \label{eq:Guentherderivative}
   ||\guentGrad \mathbf{d}||^2=||\gradS \mathbf{d}||^2 + \langle\boldsymbol{\nu} \otimes\boldsymbol{\curvTensor}\mathbf{d},\boldsymbol{\nu} \otimes \boldsymbol{\curvTensor}\mathbf{d}\rangle 
\end{equation}
where $\boldsymbol{\curvTensor}=-\tangGrad \boldsymbol{\nu}$ denotes the shape operator and $\langle \cdot, \cdot\rangle$ the scalar product on $\surf$  \cite{dziuk2013finite}. Thereby $\guentGrad$ denotes the Guenther derivative and $\tangGrad$ the surface tangential gradient, see SI. There are various physical implications resulting from the choice of derivative, see \cite{Nestler_SM_2020} for an overview. Relevant to our case is only the alignment of $\mathbf{d}$ with principal curvature directions resulting from the second summand in eq. \eqref{eq:Guentherderivative}. This coupling term has been added in an ad hoc manner in \cite{Bell_PRL_2022} to account for linear curvature contributions in surface active nematodynamics. We consider it in the phase field context for each cell and define the extrinsic curvature part of the free energy by
\begin{equation} \label{eq:FEC}
    \FLC=\LCF \sum\limits_{i=1}^N \surfint{\langle\normal \otimes\curvTensor\gradS \func_i,\normal \otimes \curvTensor\gradS \func_i \rangle },
\end{equation}
where the parameter $\LCF$ determines the preferred direction and strength of this geometric coupling. We furthermore use that the integral mean of $\gradS \func_i$ is orthogonal to the elongation of the cell and is thus related to $\mathbf{d}_i$. While $\FLC$ can become negative, the area conservation of $\func_i$ and $\FCA$ guarantee a well-posed problem within reasonable parameter settings. Figure \ref{fig_ext_curv} shows the effect of $\FCA$ and $\FLC$ on a single cell on different geometries if $v_0 = 0$. While the shape is independent on position in flat space and on cylinders, both having zero Gaussian curvature ($K = 0$) and being ruled surfaces, the shape depends on position on the torus. Here, $\FCA$ can be reduced by moving the cell towards regions of maximal $K$. $\FLC$ with $\LCF > 0 \; (< 0)$ deforms the cell from elongation in toroidal (poloidal) direction in regions of lowest $K$, inside, to elongation in poloidal (toroidal) direction in regions of highest $K$, outside, if the absolute maximal principal curvature direction changes from inside to outside. Further details are provided in SI. However, the influence of $\FLC$ is less pronounced on toroidal surfaces as the difference between the magnitude of the principal curvature directions is smaller than on the cylindrical surfaces.

The energy component $\FIN$ accounts for interaction between cells. We define ${\psi_i = \frac{1}{2} (\func_i + 1)}$. A common way to model repulsive and attractive forces is
\begin{align*}
\FIN \!=\! \INF \sum\limits_{i=1}^N \sum\limits_{j\neq i}\surfint{ \!\underbrace{a_{rep} \psi_i^2 \psi_j^2 - a_{att} \|\gradS \psi_i\|^2 \|\gradS \psi_j\|^2}_{:= f_{IN}}\!}
\end{align*}
with interaction strength $\IN$ and coefficients $a_{rep}$ and $a_{att}$, see \cite{Peyret_BJ_2019, Loeber_SR_2015} for the corresponding form in flat space. We modify this formulation and consider the equilibrium condition $\frac{\eps}{2} \|\gradS \phi_i\|^2 \approx \frac{1}{\eps} W(\phi_i)$ resulting from the $\tanh$-profile of $\func_i$ and approximate $a_{att} \|\gradS \psi_i\|^2 \|\gradS \psi_j\|^2 \approx \tilde{a}_{att} W(\phi_i) W(\phi_j)$, with the rescaled coefficient $\tilde{a}_{att}$. This leads to the numerically more appropriate form without derivatives, where
$$f_{IN} = \tilde{a}_{rep} (\phi_i +1)^2(\phi_j+1)^2 - \tilde{\tilde{a}}_{att} (\phi_i^2 -1)^2 (\phi_j^2 -1)^2
$$
with rescaled coefficients $\tilde{a}_{rep}$ and $\tilde{\tilde{a}}_{att}$, as in \cite{Guetal_JCP_2014,Marth_IF_2016,Jain_PRE_2022,Shen_Arxiv_2022}, see SI for the resulting short-range interaction potential.

Activity is incorporated by self-propulsion defining $\mathbf{v}_i$. There are various possibilities, which differ by complexity, ranging from random motion \cite{Loewe_PRL_2020} to considering mechanochemical subcellular processes \cite{Loeber_SR_2015,Marth_JRSI_2015} and physical implications, e.g. polarity and velocity alignment and contact inhibition \cite{Vercurysse_biorXiv_2022}, see \cite{Wenzel_PRE_2021} for a comparison. Here we define $\mathbf{v}_i= \cos(\theta_i)\mathbf{e}_1^i+\sin(\theta_i)\mathbf{e}_2^i$ with the angle $\theta_i$ which is controlled by rotational noise $d\theta_i(t)=\sqrt{2D_r}dW_i(t)$ with diffusivity $D_r$ and a Wiener process $W_i$ and the local orthonormal coordinate system $(\mathbf{e}_1^i,\mathbf{e}_2^i)$ in the tangent plane of the center of mass of cell $i$. We consider an elongation model with $\mathbf{e}_1^i$ pointing in the direction of largest elongation, similar to the approach  considered in flat space in \cite{Mueller_PRL_2019}. In each time step the preferred direction of movement is set by the largest elongation with some noise centered around this orientation.  Collective motion results from the deformability of the cells \cite{Grossmanetal_NJP_2008,Menzeletal_EPL_2012}. In the current setting all cells have the same size, cell growth and division are neglected.

The problem is solved numerically using surface finite elements \cite{dziuk2013finite,nestler2019finite} and the parallelization concept introduced in \cite{Praetorius_NIC_2017}, see SI for details, which includes Refs. \cite{Bachini_ArXiv_2023,Brandneretal_SIAMJSC_2022,Bastian_CMA_2021,Vey_CVS_2007,Witkowski_ACM_2015,Praetorius_ANS_2022}. 

\begin{figure}[htb]
    \centering
    \includegraphics[width=0.49\textwidth]{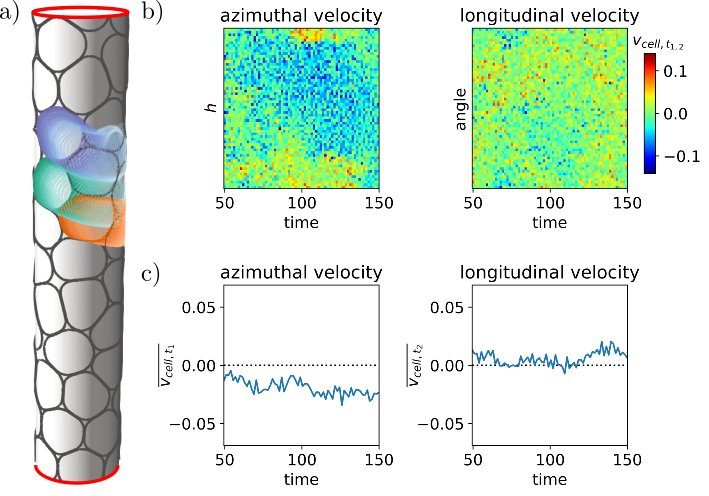}
    \caption{Evolution on a cylinder. a) Time instance of the evolution together with overlayed cell shapes ($\phi_i = 0$) and cells at previous time steps for three cells. For corresponding movie see SI. b) and c) Kymographs and graphs displaying the average velocities of the cells from a) in azimuthal and longitudinal directions as function of time. }
    \label{fig:time_evolution}
\end{figure}

We consider three cylindrical surfaces with equal surface area $|\surf|$ but different curvature and 60 equally sized cells with a packing fraction of $90\%$ placed on them with random initial direction of movement. For geometric quantities and parameters see SI. Figure \ref{fig:time_evolution} shows data for one cylinder and $\LCF > 0$, clearly indicating collective rotation, consistent with the experiments for MDCK cells in \cite{Glentis_SA_2022}.

\begin{figure*}[htb]
    \centering
    \includegraphics[width=1.0\textwidth]{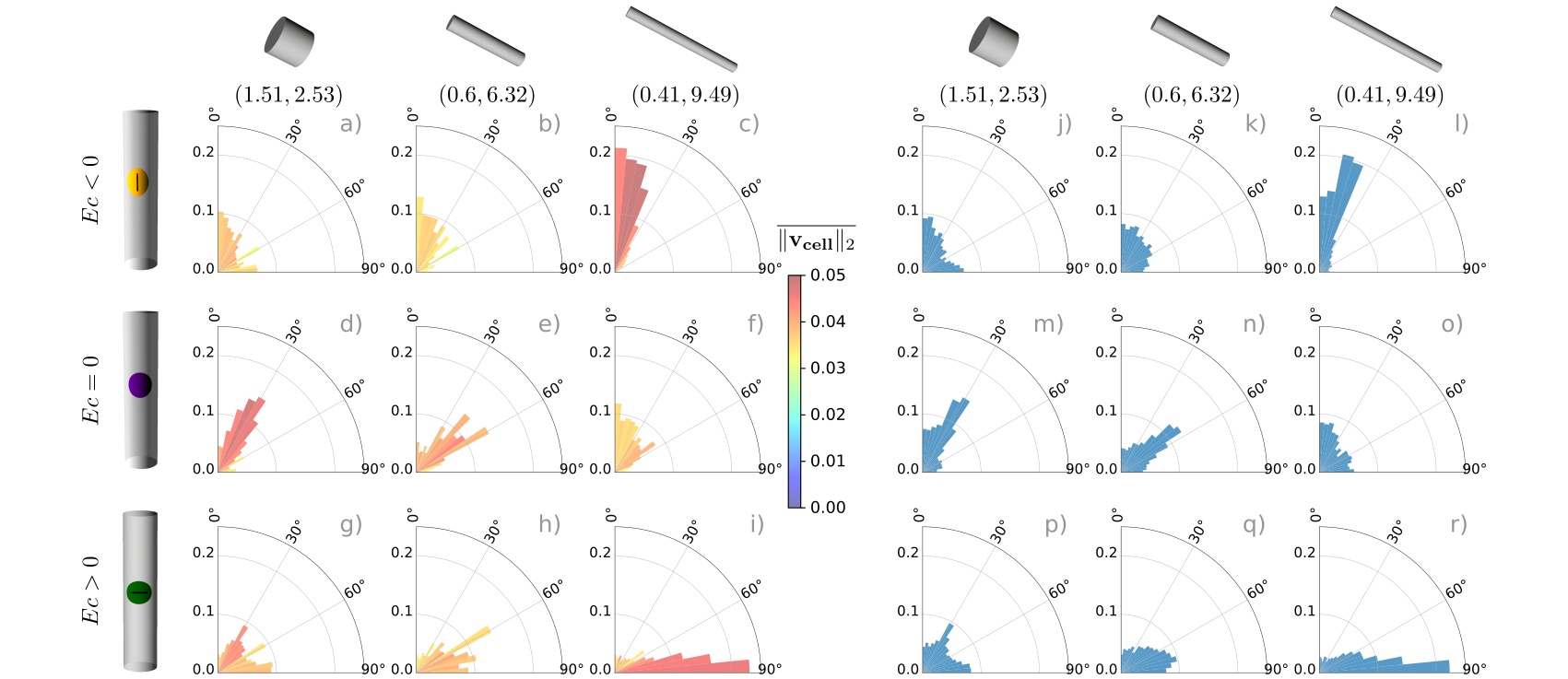} 
    \caption{Distribution of direction of motion and elongation direction on cylindrical surfaces ($r_{Cyl}, h_{Cyl}$). The angle between longitudinal direction and direction of movement or elongation direction is used as angular coordinate and the ratio of cells with this property as radial coordinate. $a)$-$i)$ Direction of movement color coded by mean velocity, $j)$-$r)$ direction of elongation. The data are averaged over time and three simulations for each configuration.}
    \label{fig_elong_cylinder}
\end{figure*}

All simulations on cylindrical surfaces are summarized in Figure \ref{fig_elong_cylinder}. We consider each cell within a time frame after an initialization phase used to randomize the cell ordering and plot the distribution of their orientation and direction of movement with respect to the angle with the longitudinal direction for three different simulations, see SI for details. The color coding corresponds to the magnitude of the averaged velocity. Without extrinsic curvature contribution ($\LCF = 0$ , see Figure \ref{fig_elong_cylinder} d) - f)) no clear trend is visible for any preferred direction of elongation or movement. The preference for $30^\circ$ and $60^\circ$ in d) and e), can be associated with a size constraint resulting from the high packing fraction. These numbers correspond to the equilibrium configuration of a hexagonal packing of circles. Analysing the neighbor distribution (not shown) confirms a more dominated hexagonal packing compared with Figure \ref{fig_elong_cylinder} f), for which the direction of movement is more distributed. For $\LCF < 0$ and $\LCF > 0$ the cells collectively elongate and move in the longitudinal and azimuthal direction, respectively. Any size constraint is overcome by the complex interaction of cells with curvature and each other. Increasing curvature enhances these effects. This is associated with stronger elongation, more pronounced movement in longitudinal or azimuthal direction and increased velocity, see Figure \ref{fig_elong_cylinder} a) - c) and g) - i). The detailed data in Figure \ref{fig:time_evolution} correspond to h). Corresponding data for a) - i) are provided in SI. While the effect of extrinsic curvature is rather small for single cells, it is enhanced in coordinated motion leading to qualitatively different behaviour. However, the enhancement of the elongation with principal curvature directions also strongly depends on the self-propulsion mechanism. Corresponding results for a random model, where $\mathbf{e}_1^i$ is chosen as the direction of the velocity vector from the last time step, which can be considered as a generalization of active Brownian particles on surfaces to deformable objects \cite{Loewe_PRL_2020}, are shown in SI. For the considered parameters this mechanism leads to a preferred elongation direction only for the cylindrical surfaces with the strongest curvature $(r_{Cyl}, h_{Cyl}) = (0.41, 9.49)$, but to no tendency for collective motion in azimuthal or longitudinal direction. 

\begin{figure*}[htb]
    \centering
    \includegraphics[width=1.0\textwidth]{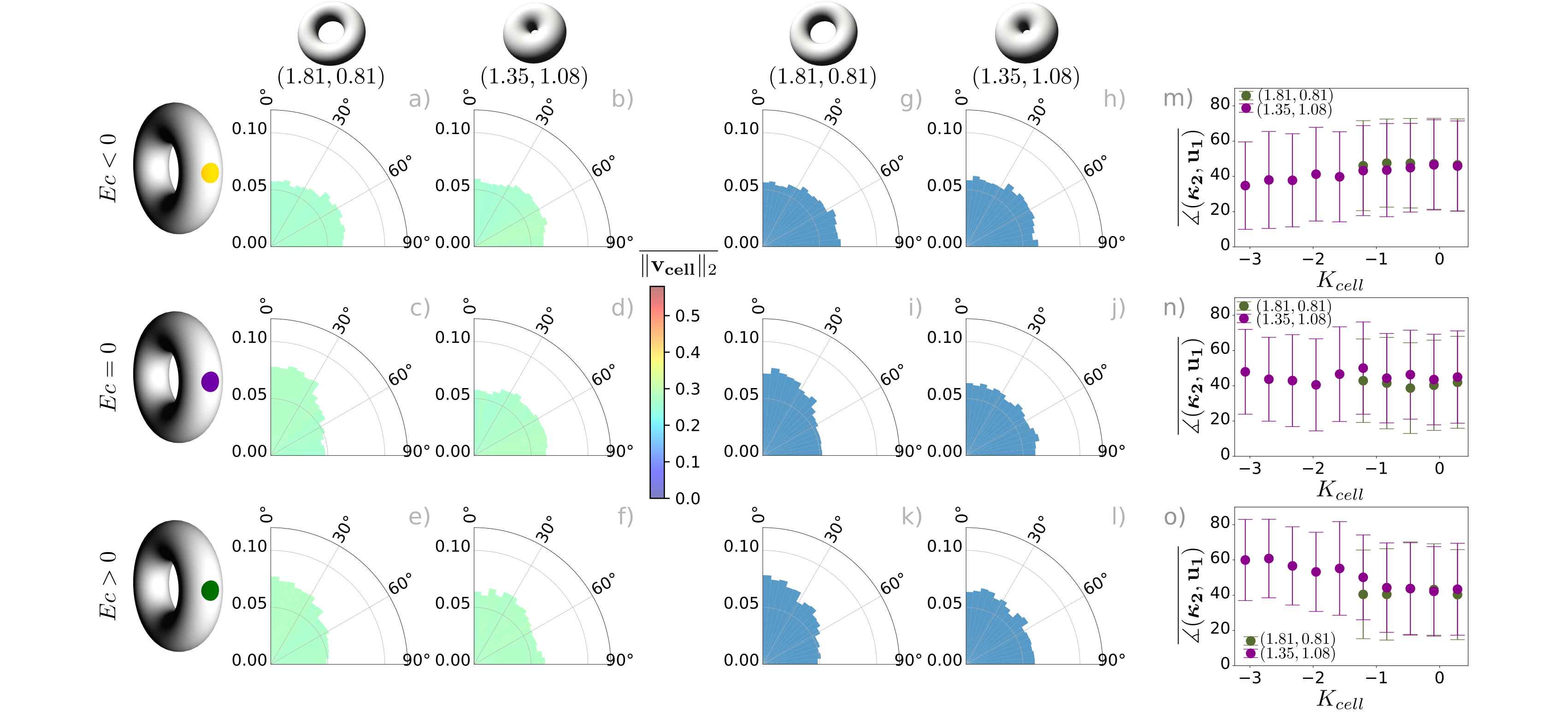} 
    \caption{Distribution of direction of motion and elongation direction on toroidal surfaces ($R_T, r_T$). The angle between poloidal direction and direction of movement or elongation direction is used as angular coordinate and the ratio of cells with this property as radial coordinate. $a)$-$f)$ Direction of movement color coded by mean velocity, $g)$-$l)$ direction of elongation. The data are averaged over time and three simulations for each configuration. $m)$-$o)$ Angle of elongation direction as function of Gaussian curvature averaged over the area of the cell $K_{cell}$.}
    \label{fig_elong_torus}
\end{figure*}

On toroidal surfaces curvature varies along the poloidal direction, see SI. As seen for a single cell, this has consequences for cell shape and position. 
We consider the same setting on two toroidal surfaces of equal area with 144 cells. Figure \ref{fig_elong_torus} summarizes the results. As in Figure \ref{fig_elong_cylinder} we plot the distribution of the direction of movement and the elongation direction. The angle is with respect to the poloidal direction.
On both surfaces the toroidal direction is the preferred direction of movement and elongation for $\LCF < 0$, see Figure \ref{fig_elong_torus} a),b),g),h) and the poloidal direction is the preferred direction of movement and elongation for ${\LCF > 0}$, see Figure \ref{fig_elong_torus} e),f),k),l). These tendencies are more pronounced for the torus with $(R_T,r_t)=(1.81,0.81)$. Here the maximal absolute principal curvature direction is always the poloidal direction, while for the torus with $(R_T,r_T)=(1.35,1.08)$ it varies from the toroidal direction (inside) to the poloidal direction (outside). This change in direction leads to preferred elongation and movement directions depending on position. In both cases varying Gaussian curvature impedes the emergence of collective motion. The difference in magnitude between the principal curvature values is smaller if compared with the cylindrical surfaces. In addition collective movement in poloidal direction is simply restricted by the geometry. These effects can explain the observed behaviour of a preferred direction of movement and elongation, but no collective motion on the torus.

We next consider the elongation as a function of $K$, see Figure \ref{fig_elong_torus} m), n), o). While we don't see an effect for the torus with $(R_T,r_T)=(1.81,0.81)$, for the torus with $(R_T,r_T)=(1.35,1.08)$ there is a slight dependency of the elongation direction on $K_{cell}$ for $\LCF < 0$ and $\LCF > 0$.  For $\LCF > 0$ it agrees, at least qualitatively, with measurements for MDCK cells on toroidal surfaces within the region of negative $K$ \cite[][Figure 2 F]{Yu_Biomaterials_2021}. Quantitative differences can be associated with significantly different numbers of cells, different measurement techniques and possible influences of the considered geometry in \cite{Yu_Biomaterials_2021}. The direction of movement has a less pronounced dependency on $K$, see SI. 

Incorporating extrinsic curvature contributions into a cell-based surface multi-phase field model allows to effectively resolve the three-dimensional nature of epithelial layers, e.g., the difference between the apical and basal side. This reveals essential effects of curvature on single cells and their collective motion. The alignment of cells with principal curvature directions leads under appropriate propulsion mechanisms and cell-cell interactions to collective motion on specific geometries. On cylindrical surfaces this can lead to long-term changes from a quiescent state to spontaneous collective rotation, as observed in vitro for MDCK cells \cite{Glentis_SA_2022}. Cylindrical surfaces are not only special mathematical objects, they are representative of many epithelial tissues, such as tubular vessels, ranging from small capillaries to large arteries, tubular glands, and ducts \cite{Bade_BJ_2018,Callens_BM_2020}. 
On more general surfaces with varying $K$ the geometric effect on the collective behaviour is a competition of intrinsic and extrinsic curvature contributions. Both couplings vastly increase the range of tissue parameters to control the flow of the epithelial layer. 
Combining this with shape changes induced by these tangential flows, as considered in coarse-grained models for fluid deformable surfaces \cite{torres2019modelling,reuther2020numerical,krause2023numerical}, has the potential to transform our understanding of morphogenesis.

{\bf Data availability:} Data and simulation code are available upon reasonable request.

{\bf Acknowledgments:} This work was funded by DFG within FOR3013. We further acknowledge computing resources at FZ Jülich under grant PFAMDIS and at ZIH under grant WIR.

\putbib[library_curvature]
\end{bibunit}
\clearpage
\newpage

\onecolumngrid

\begin{center}
  \textbf{\large \hspace{5pt} SUPPLEMENTAL MATERIAL \\ \vspace{0.2cm} Coordinated motion of epithelial layers on curved surfaces}\\[.2cm]
  L. Happel,$^{1}$A. Voigt,$^{1,2,3}$\\[.1cm]
  {\itshape \small
  ${}^1$Institute  of Scientific Computing,  TU  Dresden,  01062  Dresden,  Germany
  \\ 
  ${}^2$Center for Systems Biology Dresden, Pfotenhauerstr. 108, 01307 Dresden, Germany
 \\
  ${}^3$Cluster of Excellence, Physics of Life, TU Dresden, Arnoldstr. 18, 01307 Dresden, Germany
 
 }
 
\vspace{0.5cm}
\end{center}
\setcounter{equation}{0}
\setcounter{figure}{0}
\setcounter{table}{0}
\setcounter{page}{1}
\renewcommand{\thesection}{S\arabic{section}}
\renewcommand{\theequation}{S\arabic{equation}}
\renewcommand{\thefigure}{S\arabic{figure}}
\renewcommand{\bibnumfmt}[1]{[S#1]}
\renewcommand{\citenumfont}[1]{S#1}
\begin{bibunit}[plainnat]
\section{Geometry}

\begin{figure*}[htb]
    \centering
    \includegraphics[width=0.5\textwidth]{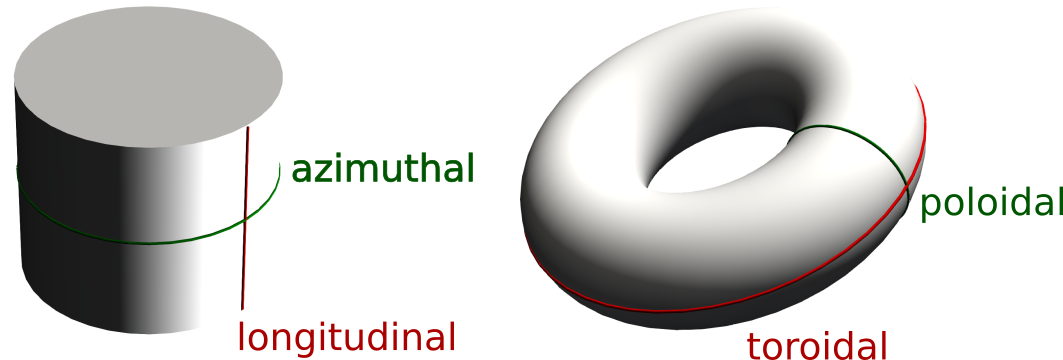} 
    \caption{Principal directions of curvature. a) for cylindrical and b) for toroidal surfaces. }
    \label{fig_notation}
\end{figure*}

\begin{figure*}[htb]
    \centering
    \includegraphics[width=0.8\textwidth]{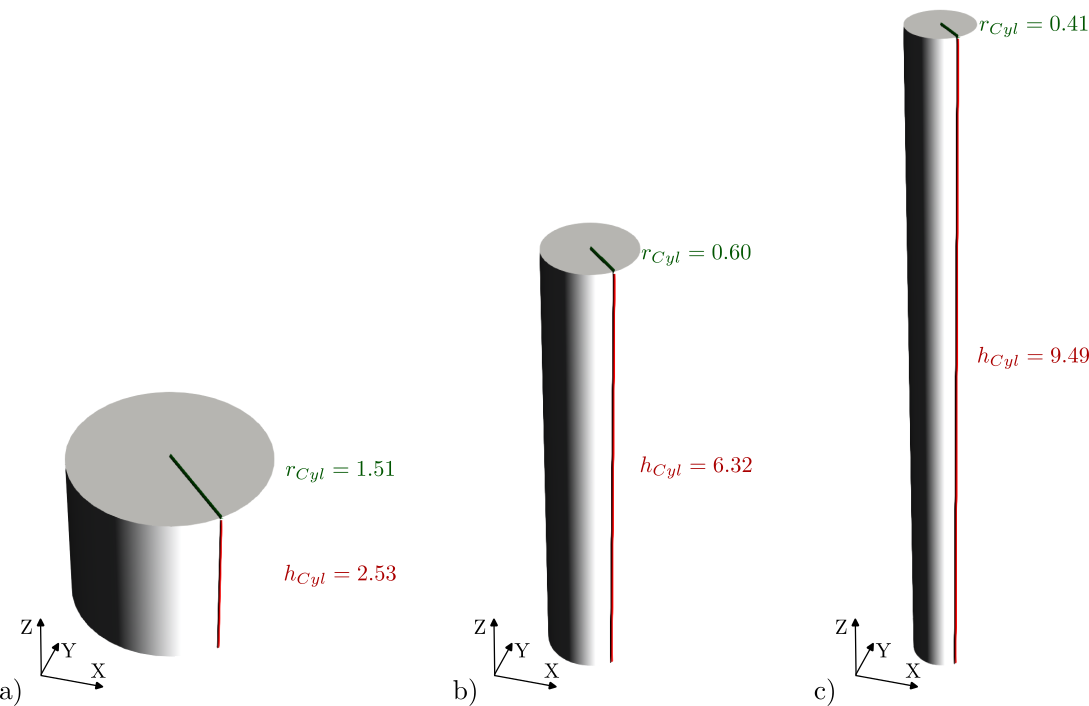} 
    \caption{Cylindrical surfaces used for the simulations with corresponding values $(r_{Cyl}, h_{Cyl})$.}
    \label{fig_geom_cylinder}
\end{figure*}

\begin{figure*}[htb]
    \centering
    \includegraphics[width=0.95\textwidth]{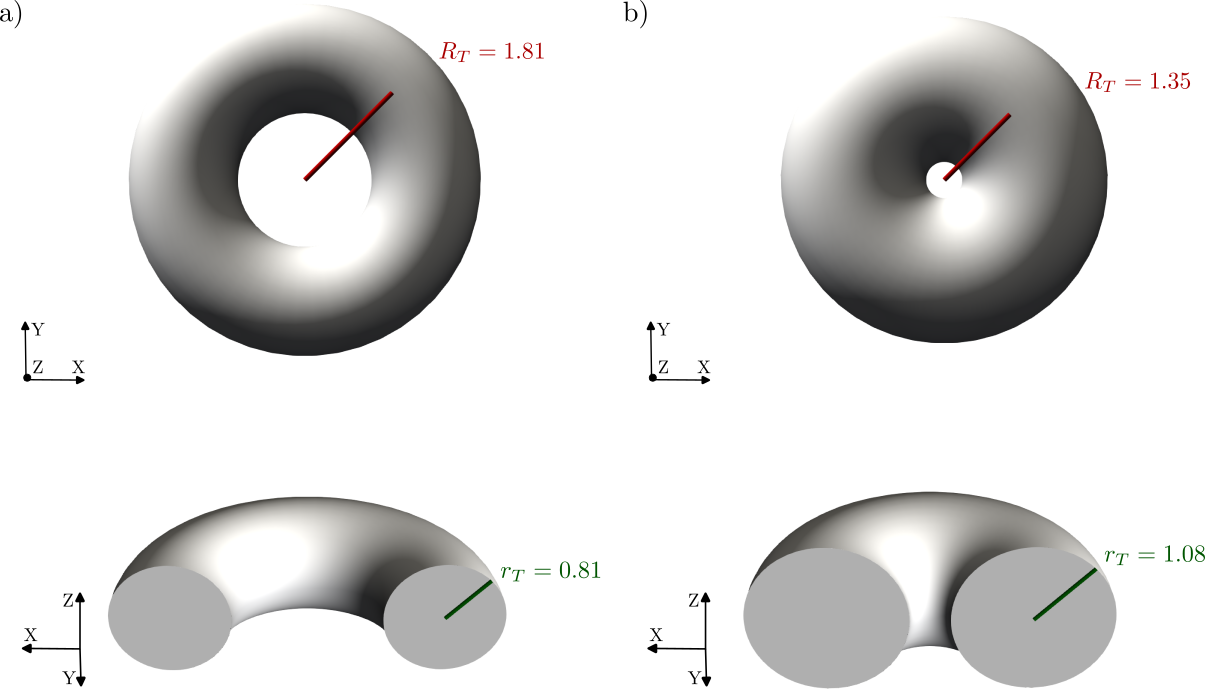} 
    \caption{Toroidal surfaces used for the simulations with corresponding values $(R_{T}, r_{T})$.}
    \label{fig_geom_torus}
\end{figure*}

We consider two prototypical geometries, cylindrical and toroidal surfaces. With periodic boundary conditions these geometries are topologically equivalent to a flat torus but differ in their geometric properties. Figure \ref{fig_notation} shows the principal directions of curvature for these geometries.
Cylindrical surfaces are characterized by their radius $r_{Cyl}$ and height $h_{Cyl}$, see Figure \ref{fig_geom_cylinder}. They are labeled as $(r_{Cyl},h_{Cyl})$. The principal curvatures are $k_1 = 1/r_{Cyl}$ and $k_2 = 0$, corresponding to the azimuthal and longitudinal direction, respectively. The three cylindrical shapes have the same area $|\surf| = 2 \pi r_{Cyl} h_{Cyl}$.
Toroidal surfaces are characterized by two radii $R_T$ and $r_T$, see Figure \ref{fig_geom_torus}. They are labeled as $(R_T,r_T)$. The principal curvatures are $k_1 = (\sqrt{x_1^2 + x_2^2} - R_T)/r_T$ in toroidal direction and $k_2 = 1/r_T$ in poloidal direction. The two toroidal shapes have the same area $|\surf| = 4 \pi^2R_Tr_T$. However, they strongly differ with respect to the Gaussian curvature $K = k_1 k_2$.

\section{Differential geometry}

Related to the surface $\surf$ we denote the outward pointing surface normal $\normal$, the shape operator (negative of the extended Weingarten map) $\curvTensor$ with $\boldsymbol{\curvTensor}=-\tangGrad \boldsymbol{\nu}$ and the surface projection $\mathbf{P}=\mathbf{I}-\normal\otimes\normal$. Let $\gradS$ be the covariant derivative. This operator is well defined for vector fields in the tangent bundle of $\surf$. For the tangential director field $\mathbf{d}$ we use the Guenther derivative, which is a component-wise tangential derivative defined as $\guentGrad \mathbf{d} = (\nabla \mathbf{d}^e)|_S \mathbf{P}$ where $\mathbf{d}^e$ is an extension of $\mathbf{d}$ constant in normal direction and $\nabla$ is the gradient of the embedding space $\R^3$. For tangential director fields $\guentGrad$ relates to the covariant derivative $\gradS$ by ${\guentGrad \mathbf{d} = \gradS \mathbf{d} +\normal \otimes\curvTensor\mathbf{d}}$, 
see \cite{Bachini_ArXiv_2023,Brandneretal_SIAMJSC_2022,krause2023numerical}. For sufficiently smooth $\mathbb{R}^3$-vector fields $\mathbf{w}$ the tangential derivative $\tangGrad$ is defined as $\tangGrad\mathbf{w}=\mathbf{P}\nabla\mathbf{w}^{e}\mathbf{P}$. Again, $\mathbf{w}^e$ is an extension of $\mathbf{w}$ constant in normal direction. For scalar fields these derivatives are identical, e.g. $\gradS \phi = \guentGrad \phi = \tangGrad \phi$.

\section{Mathematical model}

We here provide the full set of partial differential equations resulting from Eq. (1) 
\begin{equation}
    \dt \func_i + v_0(\mathbf{v}_i \cdot \gradS\func_i) = \laplaceS \frac{\delta \F}{\delta \func_i}
\end{equation}
and the free energy
\begin{equation}
    \F = \FCA + \FLC + \FIN.
\end{equation}
We consider $\frac{\delta \F}{\delta \phi_i} = \frac{\delta \FCA}{\delta \phi_i} + \frac{\delta \FLC}{\delta \phi_i} + \frac{\delta \FIN}{\delta \phi_i}$ with
\begin{align}
    \frac{\delta \FCA}{\delta \phi_i} &= \CAF \left(-\eps\gradS\cdot\left(\frac{1}{G(\func_i)} \gradS \func_i \right)+\frac{1}{G(\func_i)}\frac{1}{\eps}W'(\func_i)+\left(\frac{1}{G(\func_i)}\right)'\left(\frac{\eps}{2}\|\gradS\func_i\|^2 + \frac{1}{\eps}W(\func_i)\right) \right)\\
    \frac{\delta \FLC}{\delta \phi_i} &= -2 \,\LCF\, \gradS\cdot \left(\curvTensor^2\gradS\func_i\right)\\
    \frac{\delta \FIN}{\delta \phi_i} &= 2 \INF \sum\limits_{j\neq i} \left( 2\tilde{a}_{rep}(\func_i+1)(\func_j+1)^2-4\tilde{\tilde{a}}_{att}(\func_i(\func_i^2-1))(\func_j^2-1)^2\right).
\end{align}
The resulting system is a set of $N$ (number of cells) coupled 4th order surface partial differential equations.

\section{Numerical issues}
The resulting system of surface partial differential equations is solved by surface finite elements \cite{dziuk2013finite, nestler2019finite} within the toolbox AMDiS \cite{Vey_CVS_2007, Witkowski_ACM_2015} which was recently integrated into the DUNE framework \cite{Bastian_CMA_2021}. For the surface discretization an analytic grid function from DUNE-CurvedGrid \cite{Praetorius_ANS_2022} is used, which gives access to analytic formulas for the projection $\mathbf{P}$, surface normal $\boldsymbol{\nu}$ and the shape operator $\curvTensor$. An accurate representation of the surface is crucial as $\FLC$ has been shown to be sensitive to surface discretization errors.

Each cell, represented by the phase field variable $\phi_i$, is considered on its own core and has its own mesh, which is adaptively refined within the diffuse interface to ensure approximately 7 grid points across the interface. Dealing with $\FIN$ leads to a non-local problem and in principle requires communication between all cells and thus all cores. Due to the short-range interaction this communication can be reduced to the neighboring cells. This approach allows parallel scaling with the number of cells \cite{Praetorius_NIC_2017}, which has been demonstrated for up to 1,000 cells in flat space and carries over to the curved surface. 

We split the higher order partial differential equations for each $\phi_i$ into a system of second order partial differential equations by introducing $\mu_i=\frac{\delta \F}{\delta \func_i}$ and consider $P^2$-Lagrange elements for the unknowns $\phi_i$ and $\mu_i$. 

Discretization in time is done by finite differences using $$\dt \func_i \approx \frac{\func_i^{n+1}-\func_i^n}{\tau_n},$$ where $\tau_n$ denotes the time step size for the $n-$th time step. It is chosen to fulfill the CFL condition. In general a linear implicit-explicit scheme is used, where all linear terms treated implicitly and all nonlinear terms explicitly. However, the double-well potential $W(\phi_i)$ and the non-linear terms in $\FIN$ are linearized with one Newton-step and the de Gennes factor $G(\phi_i)$ is regularized by 
\begin{equation}
    G_\eta(\phi_i) = \sqrt{\frac{9}{4} (1 - \phi_i^2)^2 + \eta^2 \epsilon^2},
\end{equation}  
with $\eta > 0$. We further use the equilibrium $\tanh$-profile of the phase field variables $\phi_i$ to simplify $\frac{\delta \FCA}{\delta \phi_i}$, see \cite{Salvalaglio_MMAS_2021}. This results in 
\begin{align}
    \frac{\delta \FCA}{\delta \phi_i} &\approx \CAF \left(-\eps \frac{1}{G(\func_i)} \laplaceS \func_i +\frac{1}{G(\func_i)}\frac{1}{\eps}W'(\func_i)\right).
\end{align}
The resulting linear system in each time step is solved by the direct solver UMFPACK.

To extract the elongations of the cells the eigenvalues and eigenvectors of the surface Q-tensors $\mathbf{q}_i$ have to be computed. We consider the tangent plane at the center of mass $\mathbf{x}_{CM}$ of the cell and specify an orthonormal basis for this plane, to be precise we use the (normalized) principal curvature directions at the center of mass of the cell as  $\mathbf{t_1}(\mathbf{x}_{CM})$ and $\mathbf{t_2}(\mathbf{x}_{CM})$. One way to define the surface Q-tensor $\mathbf{q}_i$ would be to project the cell on this tangent plane and use
the definition of the Q-tensor in flat space \cite{Mueller_PRL_2019,Wenzel_PRE_2021} to compute the elongation. Numerical experiments using this approach did not lead to the desired results as such a projection induces a bias of the elongation in the direction of absolute minimal principal curvature. In order to avoid such bias we define the surface Q-tensor $\mathbf{q}_i$ on the surface $\surf$ by integration over the cell boundary using the local Darboux frame with $\mathbf{t_1}(\mathbf{x})$ and $\mathbf{t_2}(\mathbf{x})$ orthonormal vectors in the tangent bundle and $\normal(\mathbf{x})$ the outward-pointing normal to the surface at $\mathbf{x} \in {\cal{S}}$, where $\mathbf{t_1}(\mathbf{x})$ and $\mathbf{t_2}(\mathbf{x})$ are related to  $\mathbf{t_1}(\mathbf{x}_{CM})$ and $\mathbf{t_2}(\mathbf{x}_{CM})$ by parallel transport. On the cylinder the resulting orthonormal vectors $\mathbf{t_1}(\mathbf{x})$ and $\mathbf{t_2}(\mathbf{x})$ are the corresponding (normalized) principal curvature directions at $\mathbf{x}$. On the torus calculating the parallel transported basis requires more work. We choose the global parametrization of the surface $ 
F: \Omega \subset \mathbb{R}^2 \longrightarrow \surf$ as 
$$
F(\varphi, \theta)=\begin{pmatrix}
    (R_T+r_T\cos(\theta))\cos(\varphi) \\
    (R_T+r_T\cos(\theta))\sin(\varphi) \\
    r_T\sin(\theta)
\end{pmatrix},
$$ 
with $\varphi \in [-\pi,\pi), \theta \in [-\pi,\pi)$ and $\Omega$ a subset of the parameter space and solve the geodesic equation as a boundary value problem in the parameter space, where the center of mass $\mathbf{x}_{CM}$ corresponds to the virtual time $\mathsf{t}=0.0$ and the point in the interface $\mathbf{x}$ to the virtual time $\mathsf{t}=1.0$. For the geodesic equation we consider a smooth map $w: [0.0,1.0] \longrightarrow \Omega$. The path $\gamma=F \circ w$ is a geodesic on $\surf$, if 
$$
\ddot{w}^\nu=-\Gamma^\nu_{\alpha \beta}\dot{w}^\alpha\dot{w}^\beta ,
$$
with $\dot{}$ the virtual time derivative, $\nu,\alpha,\beta$ indices and $\Gamma^\nu_{\alpha \beta}$ the Christoffel symbols. The standard summation convention is used. For the torus this equation simplifies as only three Christoffel symbols are not zero, to be precise $$\Gamma^\varphi_{\varphi \theta}= -\frac{r_T\sin(\theta)}{R_T+r_T\cos(\theta)},\quad \Gamma^\varphi_{\theta\varphi}= -\frac{r_T\sin(\theta)}{R_T+r_T\cos(\theta)}, \quad \Gamma^\theta_{\varphi \varphi}=\frac{1}{r_T}(R_T+r_T\cos(\theta))\sin(\theta).
$$
We solve the resulting differential equation on a one-dimensional grid using a Picard iteration to treat the non-linearity. As initial solution we use a linear interpolation between the two boundary values in coordinates in the parameter space. We thereby modify the coordinates using the periodicity to ensure that the initial solution is a path inside the cell.
To obtain the parallel transport from the solution of the geodesic equation we use that the tangent of the geodesic is always parallel to itself and that angles between vectors are preserved along the geodesic. With this we can calculate $\mathbf{t_1}(\mathbf{x})$ and $\mathbf{t_2}(\mathbf{x})$ as the parallel transport of  $\mathbf{t_1}(\mathbf{x}_{CM})$ and $\mathbf{t_2}(\mathbf{x}_{CM})$ along this geodesic. This allows to define the surface Q-tensors $\mathbf{q}_i$ as 
 \begin{equation*}
    \mathbf{q}_i=\begin{pmatrix} \surfint{\frac{(\partial_{\mathbf{t}_2}\func_i)^2-(\partial_{\mathbf{t}_1}\func_i)^2}{2}} & \surfint{-\partial_{\mathbf{t}_1}\func_i\partial_{\mathbf{t}_2}\func_i} \\[6pt]
    \surfint{-\partial_{\mathbf{t}_1}\func_i\partial_{\mathbf{t}_2}\func_i} & \surfint{\frac{(\partial_{\mathbf{t}_1}\func_i)^2-(\partial_{\mathbf{t}_2}\func_i)^2}{2}}
    \end{pmatrix}
\end{equation*}
and to compute their eigenvectors and eigenvalues. 

This approach should, e.g., ensure that a geodesic circle yields zero eigenvalues. For the cylinder this follows directly and is also ensured numerically, for the torus we at least approximate this property numerically and can confirm a significantly reduced bias of the elongation of the geodesic circle, if compared to the projection into the tangent plane of the center of mass. Testing this property requires to construct geodesic circles. Different numerical methods exist. One particularly popular approach is the heat method \cite{Crane_ACM_2013}. However, we choose a midpoint and solve the geodesic equation in every considerable close grid point, using the midpoint and the grid point as boundary values. Calculating the length of the resulting geodesics allows to approximate a geodesic circle by those points with the same geodesic distance to the midpoint. This leads to an increased accuracy if compared with the heat method, but still only approximates a geodesic circle. These errors can explain the obtained slightly nonzero eigenvalues.   

We are only interested in the eigenvalues and eigenvectors of $\mathbf{q}_i$. They can be calculated as follows
\begin{align*}
    \lambda_i^1 &= \sqrt{\left(\surfint{\frac12((\partial_{\mathbf{t_2}}\func_i)^2-(\partial_{\mathbf{t_1}}\func_i)^2)}\right)^2+\left(\surfint{-\partial_{\mathbf{t_1}}\func_i\partial_{\mathbf{t_2}}\func_i}\right)^2} \\ 
    \lambda_i^2 &= - \lambda_i^1
\end{align*}
and  
\begin{align}
    \mathbf{u}_i^{1} &=\frac{\surfint{\frac12((\partial_{\mathbf{t_2}}\func_i)^2-(\partial_{\mathbf{t_1}}\func_i)^2)} + \lambda_i^{1}}{\surfint{-\partial_{\mathbf{t_1}}\func_i\partial_{\mathbf{t_2}}\func_i}}\mathbf{t_1}(\mathbf{x}_{CM})+\mathbf{t_2}(\mathbf{x}_{CM}) \\
     \mathbf{u}_i^{2} &=\frac{\surfint{\frac12((\partial_{\mathbf{t_2}}\func_i)^2-(\partial_{\mathbf{t_1}}\func_i)^2)} + \lambda_i^{2}}{\surfint{-\partial_{\mathbf{t_1}}\func_i\partial_{\mathbf{t_2}}\func_i}}\mathbf{t_1}(\mathbf{x}_{CM})+\mathbf{t_2}(\mathbf{x}_{CM}).
\end{align}
All the integrals are well-defined as $\mathbf{t_1}(\mathbf{x})$ and $\mathbf{t_2}(\mathbf{x})$ are related by parallel transport to $\mathbf{t_1}(\mathbf{x}_{CM})$ and $\mathbf{t_2}(\mathbf{x}_{CM})$.
$\mathbf{u}_i^1$ is the eigenvector pointing in the direction of largest elongation and $\mathbf{u}_i^2$ is the one pointing in the direction of largest contraction of cell $i$. The director is thus defined as $\mathbf{d}_i = \mathbf{u}_i^1 / \|\mathbf{u}_i^1 \|$ and corresponds to $\mathbf{e}_1^i$ in the definition of $\mathbf{v}_i$. 

To take the periodicity of the domain into account when calculating the center of mass we follow the approach suggested in \cite{Bai_JGT_2008}.

Using that the shape operator $\curvTensor$ has an orthonormal basis out of eigenvectors and that the eigenvectors of $\curvTensor$ are the principal curvature directions $\boldsymbol{\kappa}_1, \boldsymbol{\kappa}_2$ (with the values of principal curvature $\princCurv_1,\princCurv_2$ as eigenvalues) and the surface normal $\normal$ with eigenvalue $0.0$ we can rewrite the extrinsic curvature terms as
\begin{equation} \label{ec_gen}
\langle\boldsymbol{\nu} \otimes \boldsymbol{\curvTensor}\mathbf{d}_i,\boldsymbol{\nu} \otimes \boldsymbol{\curvTensor}\mathbf{d}_i\rangle =\princCurv_1^2\langle\boldsymbol{\kappa}_1,\mathbf{d}_i\rangle^2+\princCurv_2^2\langle\boldsymbol{\kappa}_2,\mathbf{d}_i\rangle^2.
\end{equation}
For cylindrical surfaces with radius $\rCyl$ we thus obtain
\begin{equation} \label{ec_cyl}
\langle\boldsymbol{\nu} \otimes \boldsymbol{\curvTensor}\mathbf{d}_i,\boldsymbol{\nu} \otimes \boldsymbol{\curvTensor}\mathbf{d}_i\rangle =\princCurv_1^2\langle\boldsymbol{\kappa}_1,\mathbf{d}_i\rangle^2+\princCurv_2^2\langle\boldsymbol{\kappa}_2,\mathbf{d}_i\rangle^2 = \frac{1}{\rCyl^2}\langle\boldsymbol{\kappa}_1,\mathbf{d}_i\rangle^2,
\end{equation}
as $\princCurv_2 = 0$ and $\princCurv_1 = \frac{1}{\rCyl}$. In our setup for a cylinder the direction of the zero curvature is always aligned with the $z-$axis. Therefore the direction of the non-zero curvature is $\boldsymbol{\kappa}_1=\normal \times \mathbf{e}_z$. In this setting, $\FLC$ simplifies to
\begin{equation}
    \FLC=\LCF \sum\limits_{i=1}^N \surfint{\langle\normal \otimes\curvTensor\gradS \func_i,\normal \otimes \curvTensor\gradS \func_i \rangle }=\LCF \sum\limits_{i=1}^N \surfint{\frac{1}{\rCyl^2}\langle \boldsymbol{\kappa}_1, \gradS \func_i \rangle^2}.
\end{equation}
On a torus the directions and values of the principal curvature depend on the position $\mathbf{x}$. As in our setting the torus is obtained by the rotation of a circle around the $z-$axis the toroidal direction is $\boldsymbol{\kappa}_1=\begin{pmatrix} -x_2 \\x_1 \\0.0 \end{pmatrix}$. Therefore eq. \eqref{ec_gen} on the torus reads
\begin{align*} \label{ec_torus}
\langle\boldsymbol{\nu} \otimes \boldsymbol{\curvTensor}\mathbf{d}_i,\boldsymbol{\nu} \otimes \boldsymbol{\curvTensor}\mathbf{d}_i\rangle &=\princCurv_1(\mathbf{x})^2\langle\boldsymbol{\kappa}_1(\mathbf{x}),\mathbf{d}_i\rangle^2+\princCurv_2(\mathbf{x})^2\langle\boldsymbol{\kappa}_2(\mathbf{x}),\mathbf{d}_i\rangle^2 \\
&= \left( \frac{\sqrt{x_1^2+x_2^2}-R_T}{r_T} \right)^2 \Biggl \langle \frac{1}{\sqrt{x_1^2+x_2^2}} \begin{pmatrix} -x_2 \\x_1 \\0.0 \end{pmatrix},\mathbf{d}_i \Biggr \rangle^2+\frac{1}{r_T^2}\langle\normal(\mathbf{x})\times\boldsymbol{\kappa}_1(\mathbf{x}),\mathbf{d}_i\rangle^2,
\end{align*}
and we obtain
\begin{align*}
    \FLC &=\LCF \sum\limits_{i=1}^N \surfint{\langle\normal \otimes\curvTensor\gradS \func_i,\normal \otimes \curvTensor\gradS \func_i \rangle } \\
    &=\LCF \sum\limits_{i=1}^N \surfint{\left( \frac{\sqrt{x_1^2+x_2^2}-R_T}{r_T} \right)^2 \Biggl \langle \frac{1}{\sqrt{x_1^2+x_2^2}} \begin{pmatrix} -x_2 \\x_1 \\0.0 \end{pmatrix},\gradS \func_i \Biggr \rangle^2+\frac{1}{r_T^2}\langle\normal(\mathbf{x})\times\boldsymbol{\kappa}_1(\mathbf{x}),\gradS \func_i\rangle^2}.
\end{align*}
The sign of $\LCF$ determines whether the direction of largest elongation or the direction of largest contraction of the cell wants to align with the direction of largest absolute curvature. On a cylinder $\LCF > 0$ leads to an elongation of the cells in azimuthal direction (green cell in Figure 1b) in the main article) and $\LCF < 0$ leads to an elongation of the cell in longitudinal direction (yellow cell in Figure 1b) in the main article). On the torus the effect of $\LCF$ is less pronounced because extrinsic and intrinsic curvature effects compete with each other and determine the evolution and shape of the cell. The energy depends on the position of the cell and the geometry of the torus (as sketched in Figure 1c) in the main article).

Besides active driving in the direction of cell elongation we also consider a random model. This was sufficient to obtain collective rotation on the sphere \cite{Happel_EPL_2022}. However, it turns out that this mechanism does not lead to coordinated movement on a cylindrical shape, see Figure \ref{fig_random_cylinder}.  Instead of the elongation model with $\mathbf{e}_1^i$ pointing in the direction of largest elongation, we specify $\mathbf{e}_1^i$ to be the direction of movement from the previous time step. Such an approach was introduced in \cite{Loewe_PRL_2020} and can be considered as an extension of active Brownian particles to deformable objects. However, as already seen in \cite{Wenzel_PRE_2021}, where different propulsion mechanisms are compared, such an approach is not sufficient to resample basic mechanical properties.
The numerical approach can be easily adapted to consider this propulsion mechanism.

\section{Considered parameters}
Table \ref{tab:Par_vary} summarizes the parameters varied in the simulations. The remaining parameters are kept constant during all simulations and are denoted in Table \ref{tab:Par_const}. All cells are of equal size $A = 0.36$. Approximating the cells by a circle we obtain a cell radius $r_{cell} = 0.34$. All simulations correspond to an area fraction of $90\%$.
\newline
\begin{table}[h]
    \centering
    \begin{tabular}{c|c|c|c}
        $\LCF$ & $(r_{Cyl}, h_{Cyl})$ & $(R_T, r_T)$ & activity mode  \\
        \hline
    $-4.1 \cdot 10^{-6},0.0,4.1 \cdot 10^{-6}$&$(0.41,9.49),(0.60,6.32),(1.51,2.53)$& $(1.81,0.81),(1.35, 1.08)$ & elongation, random
    \end{tabular}
    \caption{Varied parameters for cylindrical and toroidal surfaces, $\LCF$ extrinsic curvature parameter, $(r_{Cyl}, h_{Cyl})$ radius and height of the cylinder, $(R_T, r_T)$ large and small radius of the torus.}
    \label{tab:Par_vary}
\end{table}

Within units of $r_{cell}$ the radius and height of the cylinder $(r_{Cyl}, h_{Cyl}) = (1.21, 27.91), (1.77, 18.59), (4.44, 7.44)$ and the large and small radius of the torus $(R_T, r_T) = (5.29, 2.38), (3.97, 3.18)$, listed in the same order as in Table \ref{tab:Par_vary}.

\begin{table}[h]
    \centering
    \begin{tabular}{c|c|c|c|c|c|c|c|c|c}
     $N_{Cyl}$ & $N_T$ & $\eps$ & $\CA$ & $\IN$ & $\tilde{a}_{rep}$& $\tilde{\tilde{a}}_{att}$ &$v_0$&$\tau_n$&$D_r$ \\   
     \hline
     $60$ & $144$ & $0.01$ & $10.0$ & $0.05$ & $0.0625$ & $0.03$ &$0.2$&$0.01$& $0.00045$   
    \end{tabular}
    \caption{Parameters used in all simulations, $N_{Cyl}$ number of cells on cylinder, $N_T$ number of cells on torus, $\eps$ width of the diffuse interface in the phase field description, $\CA$ capillary number, $\IN$ interaction number, $\tilde{a}_{rep}$ strength of repulsive interaction, $\tilde{\tilde{a}}_{att}$ strength of attractive interaction, $v_0$ self-propulsion strength, $\tau_n$ time step and $D_r$ rotational diffusion parameter.}
    \label{tab:Par_const}
\end{table}

\begin{figure*}[hbt]
    \centering
    \includegraphics[width=0.53\textwidth]{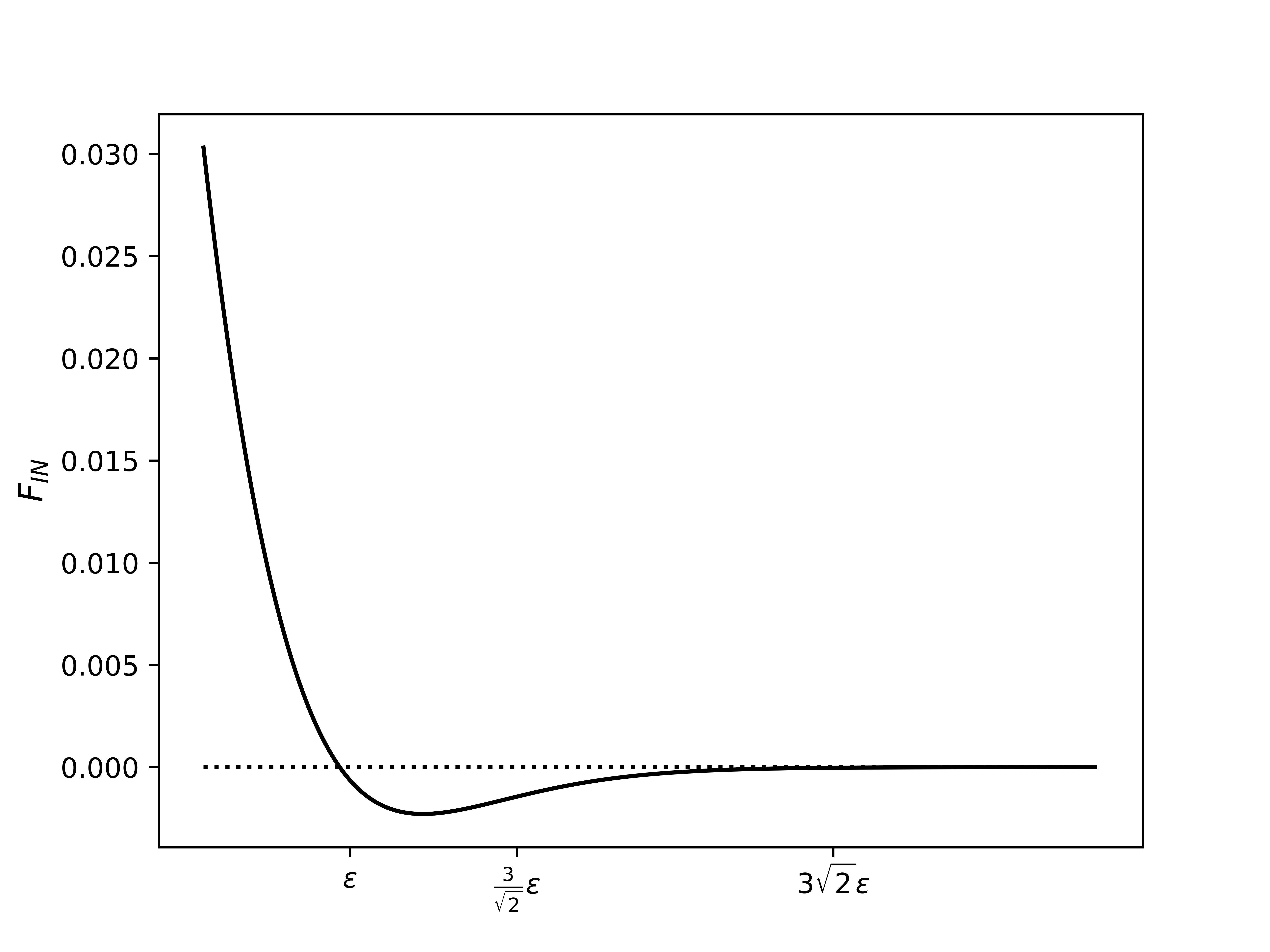} 
    \caption{Visualization of $\FIN$ for the parameters used in the simulations. Displayed is $\FIN$ between two cells in dependence of the distance of the zero contour. For a phase field with the stable phases $-1.0$ and $1.0$ the width of the interface is approximately $3\sqrt{2}\eps$ (obtained from the equilibrium tanh-profile). This point is marked explicitly.}
    \label{fig_potential}
\end{figure*}

The resulting interaction potential $\FIN$ is visualized in Figure \ref{fig_potential} as a function of distance. It is a short range potential, active only within the diffuse interface.

\section{Data analysis}

The data is evaluated after an initial time period so that our measurements are independent of the random initialization. Within this initial phase we consider a random propulsion mechanism with large $v_0$. The kymographs and averaged cell velocities over time in Figure 2 in the main article and in Figures \ref{fig_kymo_elong_cylinder} and  \ref{fig_kymo_random_cylinder} show one simulation in the time interval $[50,150]$. The statistical data on the distribution of the direction of motion and elongation direction in Figures 3 and 4 in the main article and Figure \ref{fig_random_cylinder} takes three different simulations into account. 

For the polar plots of the cylinder showing the direction of movement (see Figure $3a)$ - $3i)$ in the main article and Figure \ref{fig_random_cylinder}$a)$ - \ref{fig_random_cylinder}$i)$) the velocity vectors $\mathbf{v_{cell}}$ are calculated from the difference in $\mathbb{R}^3$ between the center of mass of the corresponding cell at time $t$ and time $t+6.0$. If the magnitude of the velocity vector is smaller than $10^{-6}$ it is not regarded, because we only want to consider velocity vectors where the movement direction surely dominates approximation errors, e.g. resulting from the approximate calculation of the center of mass. With this we have roughly $1500$ data points for each plot.

We then calculate the angle between $\mathbf{v_{cell}}$ and the longitudinal direction and compute the distribution from this. For the distribution we use $16$ bins, so we divide the interval from $0^\circ$ to $90^\circ$ into 16 equal-sized bins. For each bin we calculate the mean velocity $\overline{\|\mathbf{v_{cell}}\|}$ which is color-coded.

For the polar plots of the cylinder which show the distribution of the elongation direction (see Figure $3j)$ - $3r)$ in the main article and Figure \ref{fig_random_cylinder}$j)$ - \ref{fig_random_cylinder}$r)$) the elongation directions $\mathbf{u^1_i}$ have been calculated during run time according to eq. $(2)$. We only consider values every $6$ time units to be consistent with the evaluations for the direction of movement. Again we calculate the angle between $\mathbf{u^1_i}$ and the longitudinal direction and compute the distribution from this. For the distribution we use $16$ bins, so we divide the interval from $0^\circ$ to $90^\circ$ into 16 equal-sized bins. 

For the kymographs (Figure $2b)$ in the main article and Figures \ref{fig_kymo_elong_cylinder} and \ref{fig_kymo_random_cylinder}) the velocity in longitudinal (respectively azimuthal) direction of each cell is calculated. We calculate this from the center of mass of the corresponding cell at time $t$ and time $t+1.5$. The velocity in the azimuthal direction is calculated from the signed angle between $(x_1,x_2)$ at the two time points and $\rCyl$. For the azimuthal velocity the data points at each time point are sorted according to the height of the cell on the cylinder, i.e. according to $x_3$ of the center of mass of the cell. This makes also batch-wise rotation visible, e.g. all cells in the upper half of the cylinder rotating in one direction and all cell in the lower part of the cylinder rotating in the other direction. The longitudinal velocity is calculated from the difference of the $x_3$-coordinates at the two time points. For the longitudinal velocity the data points at each time point are sorted according to the angle between $(x_1,x_2)$ of the center of mass of the cell and the direction $(1.0,0.0)$. This makes also common movements in longitudinal direction of only a part of the cells visible. 

The calculation of the polar plots (see Figure $4a)$-$4f)$ in the main article) for the tori, which show the direction of movement, is done similarly to the calculation of the polar plots for the cylinders except for two things: First, we calculate the velocity vectors $\mathbf{v_{cell}}$ from the difference in $\mathbb{R}^3$ between the center of mass of the corresponding cell at time $t$ and time $t+1.5$, since the difference between the velocity vector in $\mathbb{R}^3$ and the velocity vector on the surface is larger for the tori than for the cylinder. Second, we calculate the angle between $\mathbf{v_{cell}}$ and the poloidal direction instead of the angle with the longitudinal direction. For this we take the poloidal direction at the center of mass of the cell at $t+0.75$.
To be consistent with the plots for the direction of movement also for the polar plots showing the direction of elongation (see Figure $4g)$-$4l)$ in the main article) the values of $\mathbf{u^1_i}$ every $1.5$ time units are used. There the angle between $\mathbf{u^1_i}$ and the poloidal direction is calculated with the poloidal direction at the center of mass at the same point in time.

The mean Gaussian curvature experienced by a cell, $K_{cell}$, is calculated at runtime using the following formulae:
\begin{equation}
    K_{cell}=\frac{\surfint{K \psi_i}}{\surfint{\psi_i}}
\end{equation}
with $\psi_i = \frac{1}{2} (\phi_i +1)$. To evaluate the direction of movement with respect to
$K_{cell}$ (see Figure $4m)$ in the main text) we compute $\mathbf{v_{cell}}$ and the angle with the poloidal direction exactly as for the polar plots. But this time the bins for the distribution are computed with respect to $K_{cell}$, where we took $K_{cell}$ at $t+0.75$. We used $10$ bins, which are equally distributed between $K_{cell}=-3.26$ and $K_{cell}=0.47$ as this where the minimal and maximal value for $K_{cell}$ encountered in the simulations. In these bins the mean angle was calculated.

For the evaluation of the elongation direction in terms of $K_{cell}$ (see Figure $4n)$ in the main text) the angles between the elongation direction and the poloidal direction are computed exactly as in the polar plots and the values of $K_{cell}$ are taken from the same time step. As for Figure $4m)$ from the main text, the bins for the distribution are computed with respect to $K_{cell}$ and $10$ equally sized bins from $K_{cell}=-3.26$ to $K_{cell}=0.47$ are used. In these bins the mean angle was calculated.
The error bars in Figure $4m)$ and $4n)$ in the main article have the length of the standard deviation of the values in that particular bin.

\section{Results for a single cell}
\begin{figure*}[!htb]
    \centering
    \includegraphics[width=0.9\textwidth]{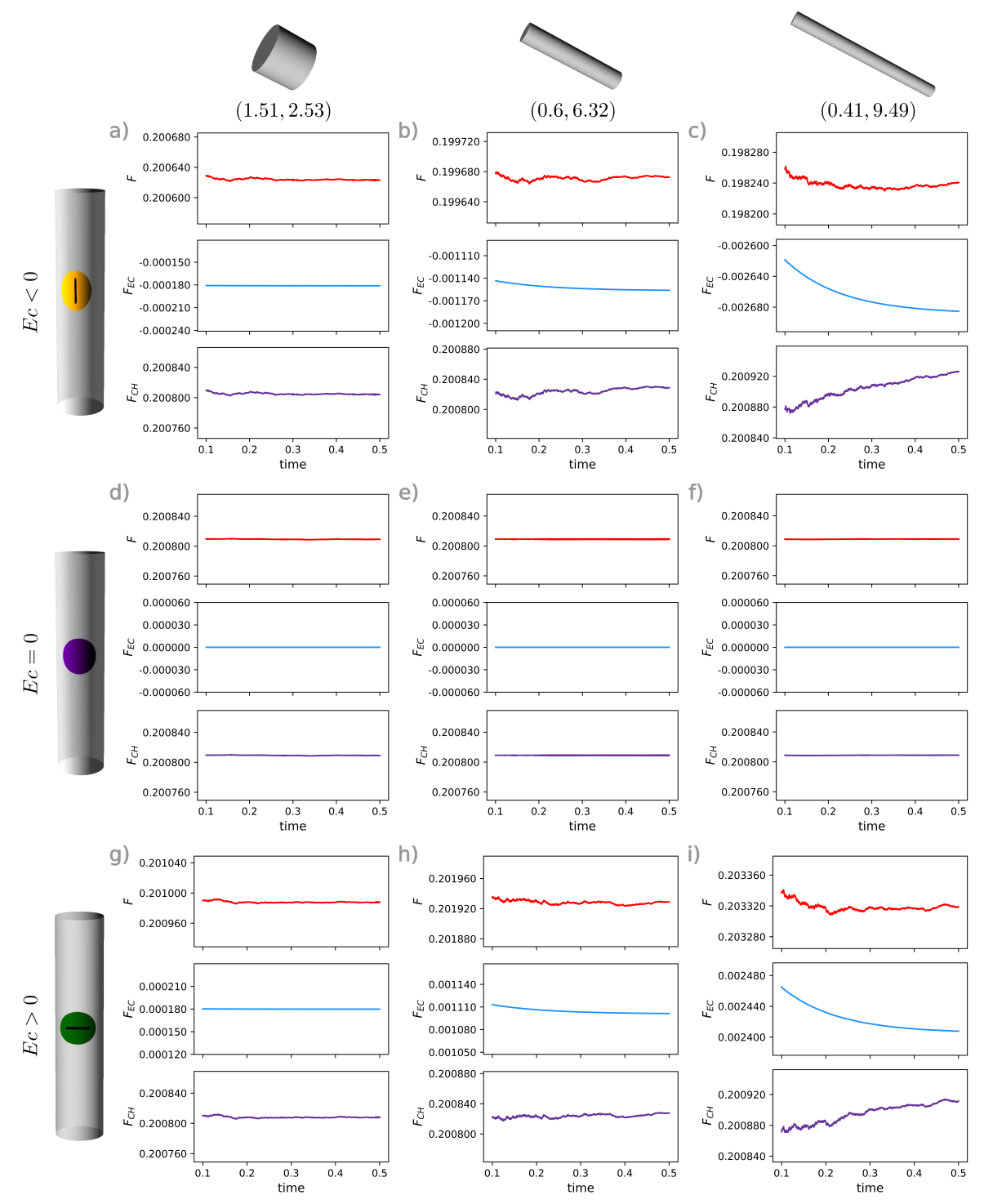} 
    \caption{Time evolution of ${F = \FCA + \FLC}$, $\FCA$ and $\FLC$ for different cylindrical surfaces and different contributions of the extrinsic curvature energy. The different rows correspond to the extrinsic curvature contribution $\LCF < 0$, $\LCF = 0$ and $\LCF > 0$ from top to bottom. The columns show the different cylindrical surfaces ($r_{Cyl}, h_{Cyl}$).}
    \label{fig_energy_cylinder}
\end{figure*}

\begin{figure*}[!htb]
    \includegraphics[width=0.66\textwidth]{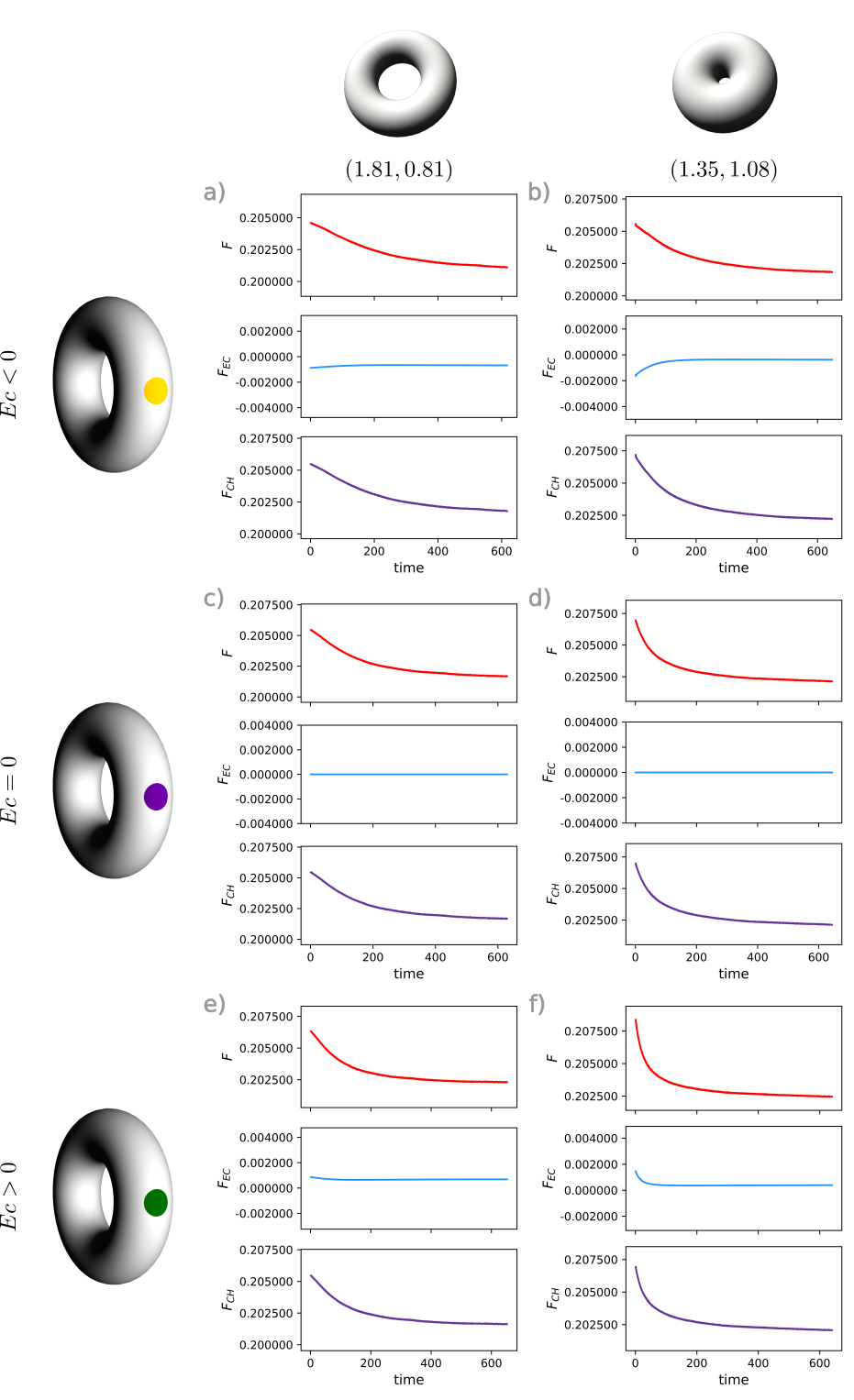}
    \caption{Time evolution of $F = \FCA + \FLC$, $\FCA$ and $\FLC$ for different toroidal surfaces and different contributions of the extrinsic curvature energy. The different rows correspond to the extrinsic curvature contribution $\LCF < 0$, $\LCF = 0$ and $\LCF > 0$ from top to bottom. The columns show the different toroidal surfaces ($R_T, r_T$).}
    \label{fig_energy_torus}
\end{figure*}

The evolution of the energy for a single cell corresponding to Figure 1 in the main article is shown in Figure \ref{fig_energy_cylinder} for the cylindrical surfaces and in Figure \ref{fig_energy_torus} for the toroidal surfaces. 

We consider the energy contributions $\FCA$ and $\FLC$ and $F=\FCA+\FLC$. We solve the evolution equation eq. (1) in the main articles with $v_0 = 0$. 

For the evaluation of one cell on the cylinder (see figure \ref{fig_energy_cylinder}) we start with a geodesic circle on the cylinder, so the optimal solution for a system with $\LCF=0.0$. This is also illustrated by the Figures \ref{fig_energy_cylinder} $d)$-$f)$, where no change in energy occurs because the solution is already optimal. If $\LCF \neq 0.0$ a small deformation of the cell can lead to a further decrease of the total energy, since the geodesic circle is no longer the optimal solution. This effect is most visible for the cylinder with the highest curvature, see Figures \ref{fig_energy_cylinder} $c)$ and \ref{fig_energy_cylinder} $i)$. 
Decreasing $\FLC$ is associated with an increase of $\FCA$. The sum of both energy contributions decreases. However, the absolute values of the energy contributions differ by three orders of magnitude. 

For the toroidal surfaces the cell is placed on the inside of the torus. This leads to a movement of the cell towards the outside of the torus. It is clearly visible that this is driven by $\FCA$ and not $\FLC$, see Figure \ref{fig_energy_torus}. The decrease of $\FCA$ is by several orders of magnitude larger than changes in $\FLC$.  

\section{Results for coordinated movement}

\begin{figure*}[!htb]
    \centering
    \includegraphics[width=0.99\textwidth]{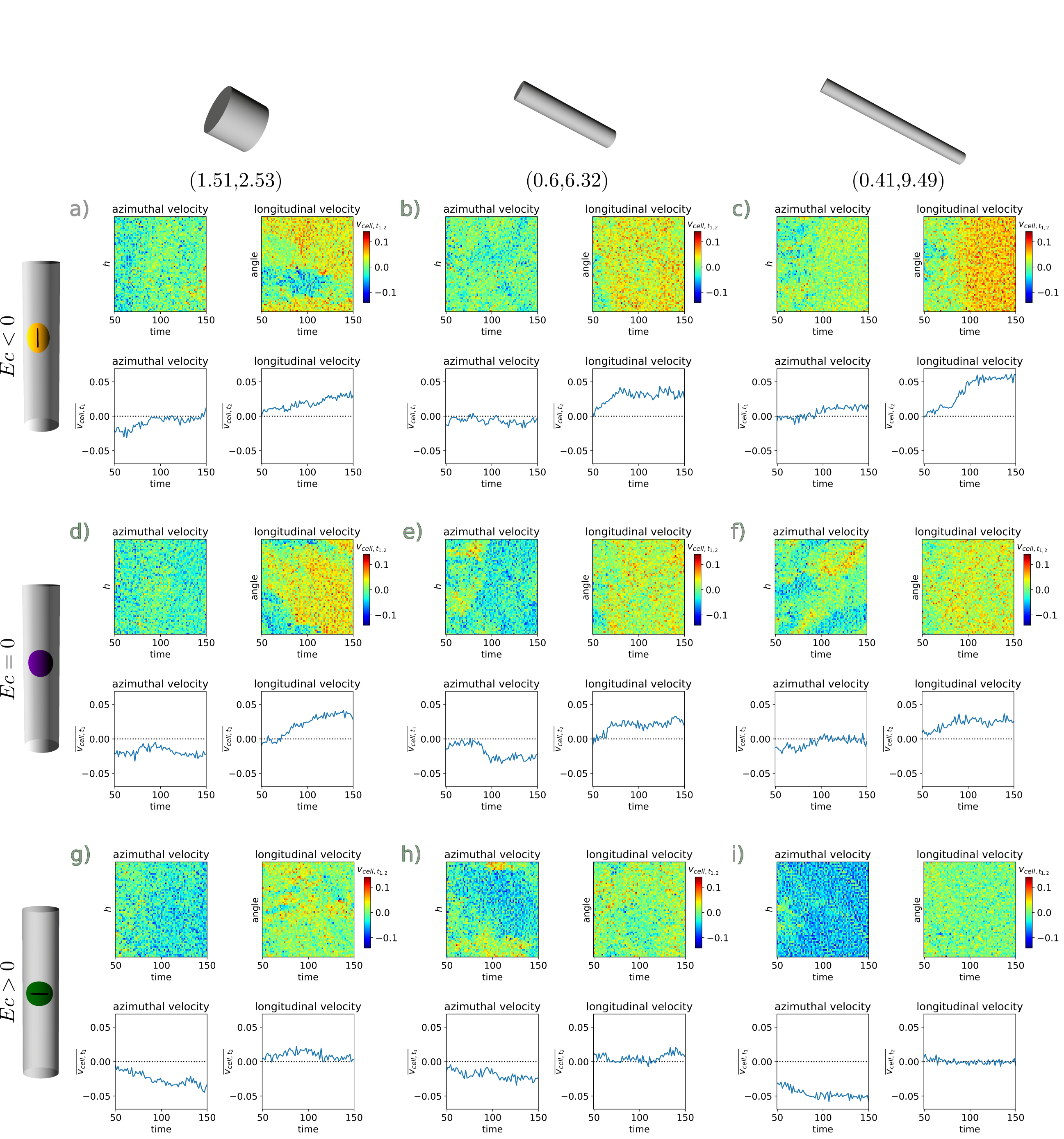} 
    \caption{Kymographs and mean azimuthal/longitudinal velocity for the elongation model on cylindrical surfaces. The different rows correspond to the extrinsic curvature contribution $\LCF < 0$, $\LCF = 0$ and $\LCF > 0$ from top to bottom. The columns show the different cylinders ($r_{Cyl}, h_{Cyl}$) with increasing curvature from left to right.}
    \label{fig_kymo_elong_cylinder}
\end{figure*}

We provide the corresponding detailed data as shown in Figure 2 of the main article for all configurations considered in Figure 3 of the main article, see Figure \ref{fig_kymo_elong_cylinder}. The results confirm the argumentation in the main article.

\begin{figure*}[!htb]
    \centering
    \includegraphics[width=1.0\textwidth]{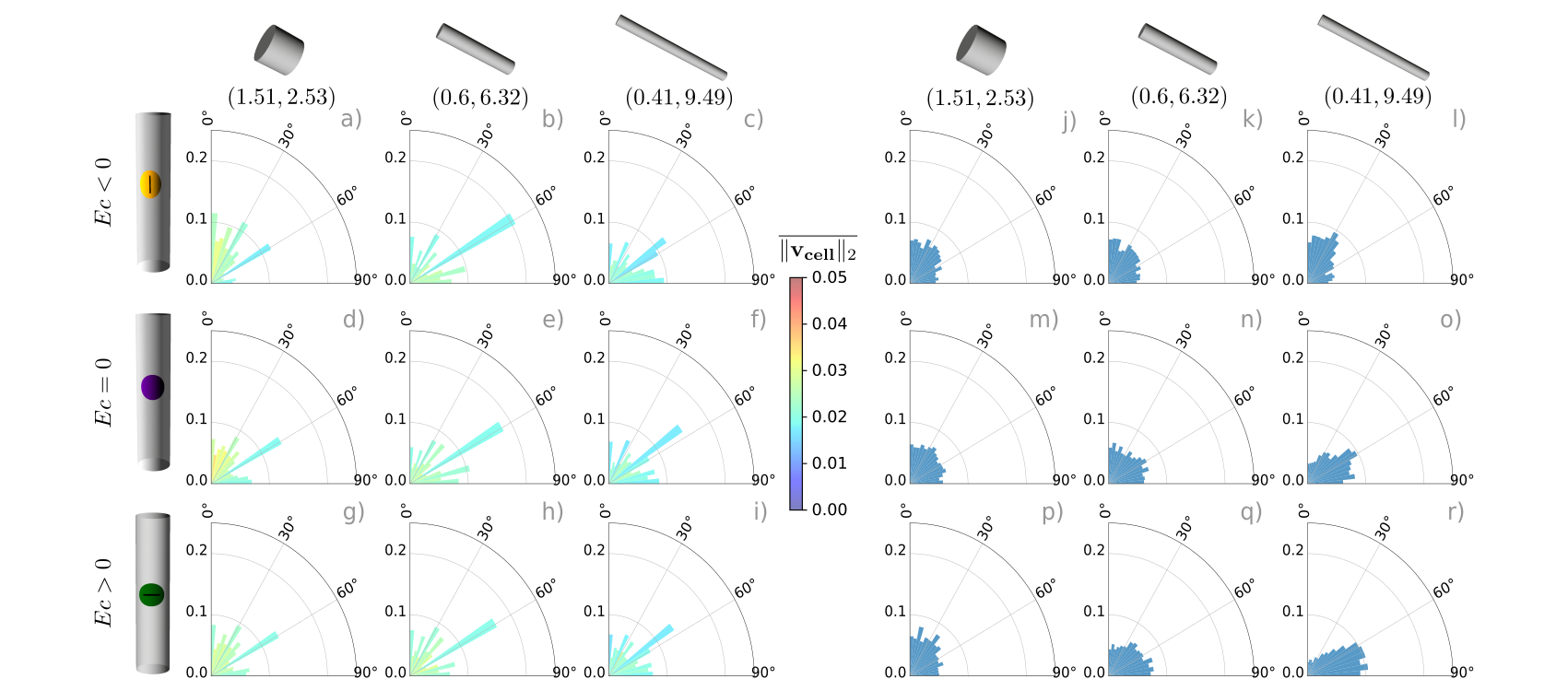} 
    \caption{Distribution of direction of motion (left) and elongation direction (right) for the random model on three cylindrical surfaces. Rows correspond to the extrinsic curvature contribution $\LCF < 0$, $\LCF = 0$ and $\LCF > 0$ from top to bottom. Columns show different cylinders ($r_{Cyl}, h_{Cyl}$) with increasing curvature from left to right. The angle between longitudinal direction and the direction of movement or elongation direction is used as angular coordinate and the ratio of cells with this property as radial coordinate. $a)$-$i)$ Direction of movement color coded by mean velocity, $j)$-$r)$ direction of elongation. The data are averaged over time and three simulations for each configuration.}
    \label{fig_random_cylinder}
\end{figure*}

\begin{figure*}[!htb]
    \centering
    \includegraphics[width=0.99\textwidth]{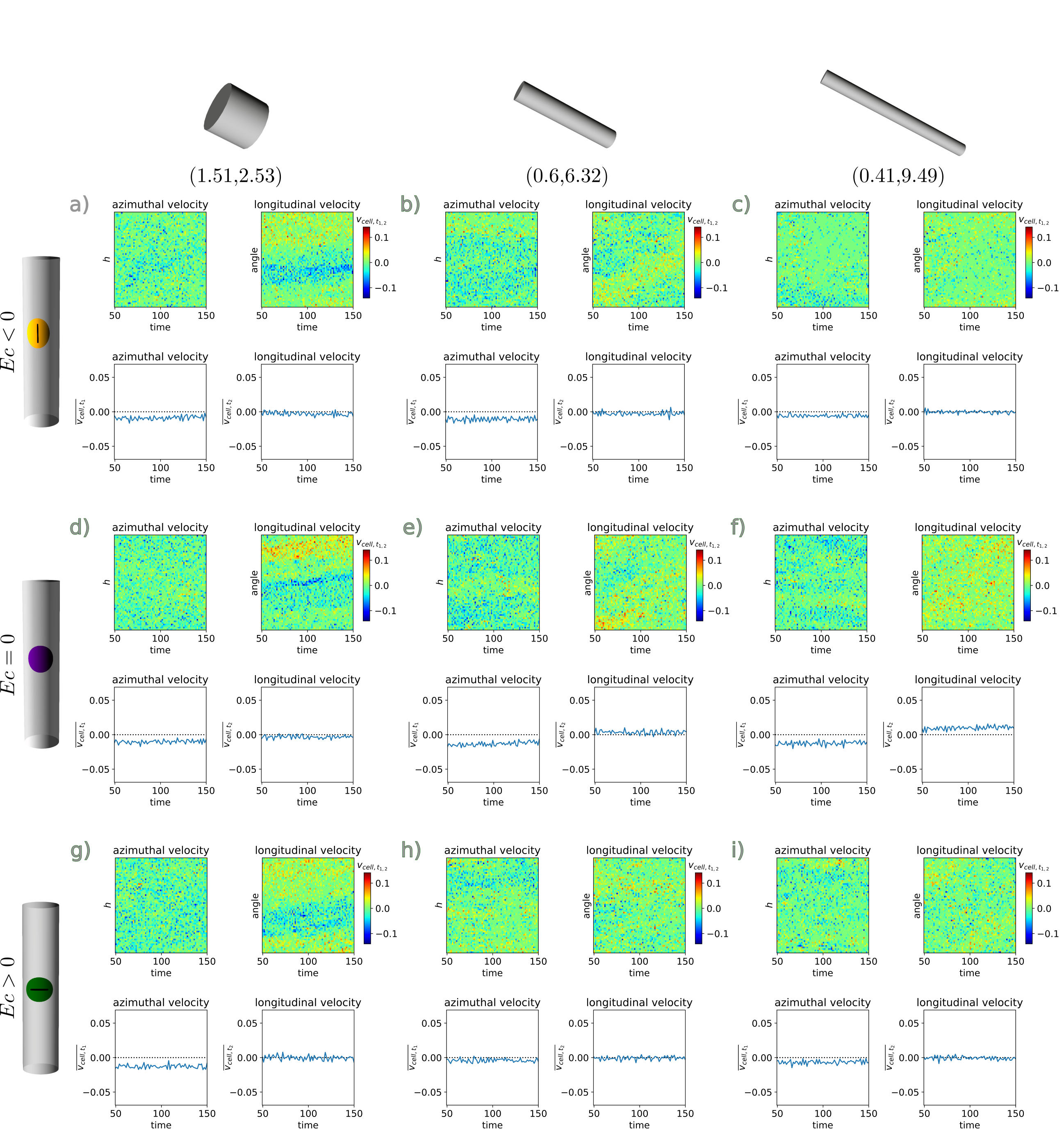} 
    \caption{Kymographs and mean azimuthal/longitudinal velocity for the random model on cylindrical surfaces. The different rows correspond to the extrinsic curvature contribution $\LCF < 0$, $\LCF = 0$ and $\LCF > 0$ from top to bottom. The columns show the different cylinders ($r_{Cyl}, h_{Cyl}$) with increasing curvature from left to right.}
    \label{fig_kymo_random_cylinder}
\end{figure*}

In addition we provide the corresponding results for a random propulsion mechanism, see figures \ref{fig_random_cylinder} and \ref{fig_kymo_random_cylinder}. While a preferred elongation direction depending on $\LCF$ is visible for the cylinder with the largest curvature, there is no preferred direction of movement.

\begin{figure*}[!htb]
    \centering
    \includegraphics[width=0.5\textwidth]{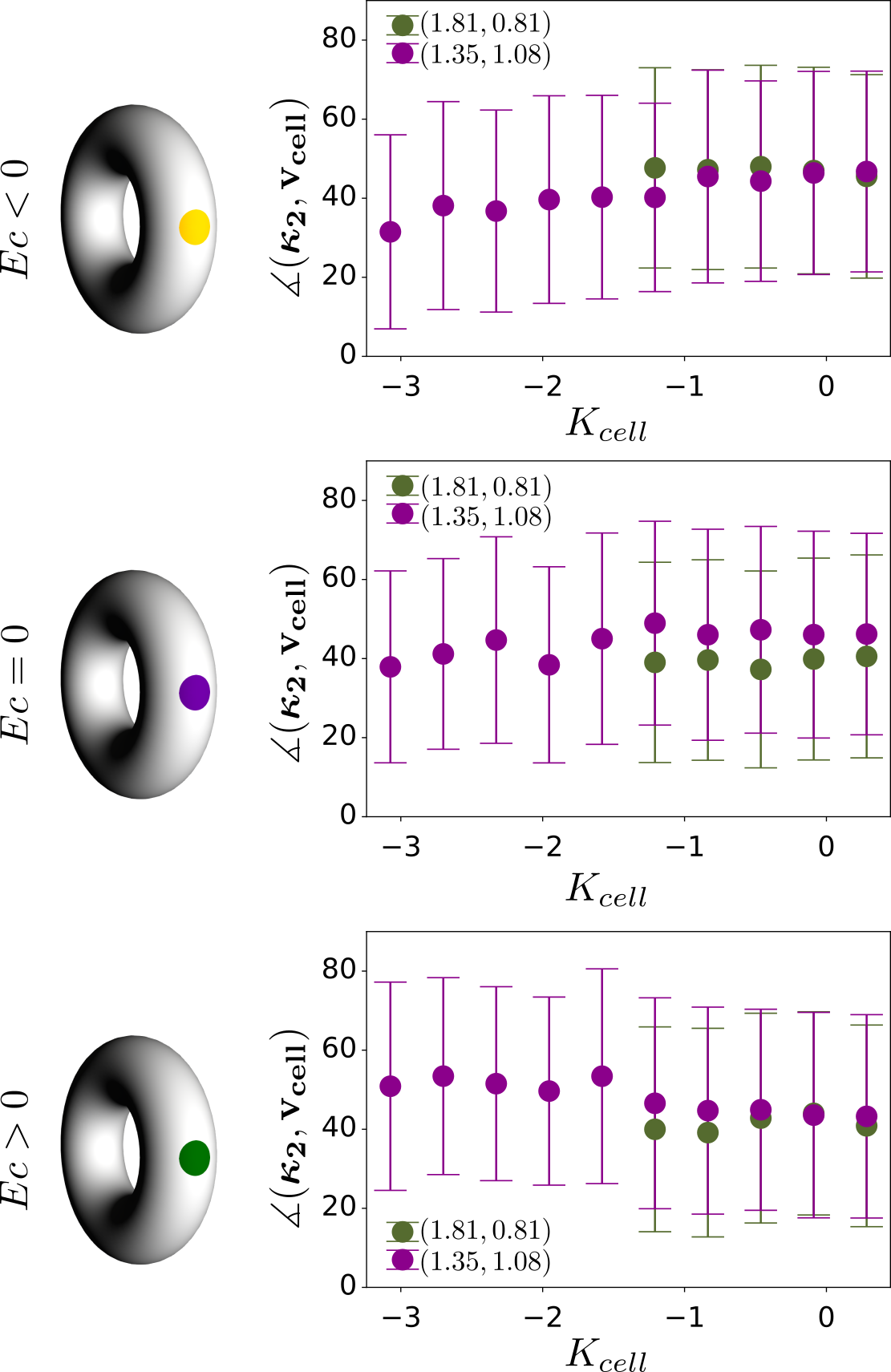} 
    \caption{Angle of direction of movement as function of Gaussian curvature of the toroidal surfaces averaged over the area of the cell $K_{cell}$. Both toroidal surfaces $(R_T,r_T)$ are shown in the same plot.}
    \label{fig_torus_movement}
\end{figure*}

Finally we show the corresponding data to Figure 4 for the dependency of the direction of movement on Gaussian curvature, see Figure \ref{fig_torus_movement}. While the elongation direction changes for $\LCF > 0$, there is no dependency visible for direction of movement, independent of $\LCF$. 

\section{Movies}

We provide movies for the considered geometries with $\LCF > 0$, modeling the behaviour of MDCK cells, corresponding to Figure 3 g), h), i) or p), q), r) in the main article for the cylindrical surfaces and Figure 4 e), f) or k), i) in the main article for the toroidal surfaces. The movies show one simulation within the time frame $[50,150]$. The cells are visualized by their zero level sets $\phi_i = 0$. 
\putbib[library_curvature]
\end{bibunit}
\end{document}